\documentclass[11pt]{article}

\usepackage[OT2,OT1]{fontenc}

\usepackage{a4wide}
\usepackage{amsmath}
\usepackage{mathrsfs}
\usepackage[T1]{fontenc}
\usepackage{mathpazo}
\usepackage{setspace}
\usepackage{amsfonts}
\usepackage{amssymb}
\usepackage{amsmath}
\usepackage{epsfig}
\usepackage{latexsym}
\usepackage{color}
\usepackage{nicefrac}
\usepackage[mathscr]{euscript}
\usepackage[latin1]{inputenc}
\usepackage{cite}
\usepackage[usenames,dvipsnames]{xcolor}
\numberwithin{equation}{section}

\author{
  \begin{minipage}{.97\linewidth}
    \vspace{1.cm}
    \begin{center}
      \begin{small}
        \textbf{P.M. Petropoulos}\footnote{marios@cpht.polytechnique.fr}
      \end{small}
    \end{center}
    \vspace{0.5cm}
    \hspace{2cm}\begin{minipage}{.7\linewidth}
     {\it \begin{footnotesize}
     \begin{center}
         Centre de Physique Th\'eorique, CNRS\footnote{Unit\'e mixte UMR7644.},
        Ecole Polytechnique, \\
        91128 Palaiseau Cedex, France
        \end{center}
     \end{footnotesize}}
    \end{minipage}
    \vspace{1.5cm}
  \end{minipage}
}

\title{\vspace{2.5cm}
 \boldmath \begin{huge}
    \textbf{Gravitational duality, topologically massive gravity and holographic fluids}
  \end{huge} \unboldmath
}

\begin{document}

\begin{titlepage}
  \maketitle
  \thispagestyle{empty}

  \vspace{-11.4cm}
  \begin{flushright}
    CPHT-RR017.0414
  \end{flushright}

  \vspace{10cm}

  \begin{center}
    \textsc{Abstract}\\
  \end{center}
Self-duality in Euclidean gravitational set ups is a tool for
finding remarkable geometries in four dimensions. From a holographic
perspective, self-duality sets an algebraic relationship between two \emph{a priori}
independent boundary data: the boundary energy--momentum tensor and the
boundary Cotton tensor. This relationship, which can be viewed as resulting from  a topological mass term for gravity boundary dynamics, survives under the Lorentzian signature
and provides a tool for generating exact bulk Einstein spaces carrying,
among others, nut charge. In turn, the holographic analysis exhibits 
perfect-fluid-like equilibrium states and the presence of non-trivial vorticity allows to show that infinite number of transport coefficients vanish.

\end{titlepage}

\onehalfspace

\tableofcontents

%\newpage

\section{Introduction}
\label{PMP-sec:intro}
Gravitational duality is known to map the curvature form of a connection onto a dual curvature  form. It allows for constructing self-dual, four-dimensional, Euclidean-signature geometries, which are in particular Ricci-flat.  Many exact solutions to Einstein's vacuum equations have been obtained in this manner, such as  Taub--NUT \cite{PMP-Taub-nut}, Eguchi--Hanson \cite{PMP-Eguchi:1978gw, PMP-Eguchi:1979yx}, or Atiyah--Hitchin \cite{PMP-Atiyah1} gravitational instantons. 

The  remarkable integrability properties underlying the above constructions have created the lore that in one way or another, integrability is related with self-duality,\footnote{Quoting Ward (1985, \cite{PMP-Ward85}): \quotation{\dots many (and perhaps all?) of the ordinary or partial differential equations that are regarded as being integrable or solvable may be obtained from the self-duality equations (or its generalizations) by reduction.}} in a general and  somewhat loose sense. In particular, this statement applies to conformal self-duality conditions, either for K\"ahler or for Einstein spaces, which have delivered many exact geometries (LeBrun, Fubini--Study, Calderbank--Pedersen, Przanowski--Tod, Tod--Hitchin, \dots \cite{PMP-GP78,  PMP-Ward80, PMP-LeBrun82, PMP-P85, PMP-P86, PMP-PS88, PMP-Pedepoon90, PMP-przanowski90, PMP-Tod90, PMP-Tod94, PMP-Hitchin95, PMP-Mas94, PMP-CP00, PMP-CP02}). 

Conformally self-dual spaces can be Einstein -- called then quaternionic. They can  be asymptotically anti-de Sitter and analyzed from a (Euclidean) holographic perspective. Hence, it is legitimate to ask (\romannumeral1) how self-duality reveals holographically \emph{i.e.} on the boundary data, (\romannumeral2) whether its underlying integrability properties extend to Lorentzian three-dimensional boundaries and allow to obtain exact bulk Einstein spaces, and
(\romannumeral3) what the physical content is for a boundary fluid emerging from such exact bulk solutions. 

The aim of these lecture notes is to provide a tentative answer to the above questions. They exhibit our present understanding of the subject, as it emerges from our works \cite{PMP-Leigh:2011au, PMP-LPP2, PMP-NewPaper,PMP-Mukhopadhyay:2013gja}. The exact reconstruction of the bulk Einstein geometry or, equivalently, the resummability of the Fefferman--Graham expansion are achieved assuming a specific relationship among the two \emph{a priori}  independent boundary data, which are the boundary metric $g_{\mu\nu}$ and the boundary momentum $F_{\mu\nu}$ interpreted as the boundary field theory  energy--momentum tensor expectation value $T_{\mu\nu}$:\footnote{The relation is $T_{\mu\nu}=\kappa F_{\mu\nu}$, given  in Eq. \eqref{PMP-FT}.}
\begin{equation}
\label{PMP-eq:sdc}
w T_{\mu\nu} + C_{\mu\nu}=0 .
\end{equation}
Here $C_{\mu\nu}$ is the Cotton--York tensor of the boundary geometry. In the Euclidean case, (anti-)self-duality corresponds precisely to the choice $w=\pm \nicefrac{3k^3}{\kappa}$ ($k$ is related to the cosmological constant, $\Lambda = -3 k^2$, and $\kappa$ to Newton's constant, $\kappa = \nicefrac{3k}{8\pi G_{\text{N}}}$). Equation \eqref{PMP-eq:sdc} appears as the natural extension of this duality -- and integrability, in the spirit of the above discussion --  requirement, irrespective of the signature of the metric, with arbitrary real $w$. This answers questions (\romannumeral1) and (\romannumeral2). Furthermore, the 
boundary condition \eqref{PMP-eq:sdc} can be recast as
\begin{equation}
\label{PMP-eq:sdcd}
\frac{\delta S}{\delta g_{\mu\nu}}
=0
\end{equation}
with 
\begin{equation}
\label{PMP-eq:sdca}
S=S_{\text{matter}} + \frac{1}{w}\int \omega_3(\gamma)  ,
\end{equation}
where $S_{\text{matter}} $ is the action of the holographic boundary matter and $\omega_3(\gamma)$ the Chern--Simons density ($\gamma$ is the boundary connection one-form). The reader will have recognized the dynamics of matter coupled to a topological mass term for gravity \cite{PMP-cs}. Exact bulk Einstein spaces  satisfying this boundary dynamics turn out to provide laboratories for probing transport properties of three-dimensional holographic fluids, and this is an important spin-off of the present analysis that will answer question~(\romannumeral3). 

\section{The ancestor of holography}
\label{PMP-sec:anc}  

We will here review some basic facts about gravitational duality and their application to the filling-in problem, which can be considered as the ancestor of holography. All this will be illustrated in the example of asymptotically AdS Schwarzschild Taub--NUT geometry.

\subsection{Curvature decomposition and self-duality}\label{PMP-sec:CDD}

The Cahen--Debever--Defrise decomposition, more commonly known as Atiyah--Hitchin--Singer
\cite{PMP-CDD, PMP-AHS},\footnote{See also \cite{PMP-Eguchi:1980jx} for a review.}
is a convenient taming of the 20 independent components of the Riemann tensor. In Cartan's formalism, these are  captured by a set of curvature two-forms ($a, b, \ldots=0, \ldots,3$)
\begin{equation}\label{PMP-curv}
 \mathcal{R}^a_{\hphantom{a}b}=\text{d} \omega^a_{\hphantom{a}b} 
 +\omega^a_{\hphantom{a}c}\wedge \omega^c_{\hphantom{c}b} = \frac{1}{2}
 R^a_{\hphantom{a}bcd} \theta^c\wedge \theta^d  ,
\end{equation}
where $\left\{\theta^a\right\}$ are a basis of the cotangent space and $   \omega^a_{\hphantom{a}b}=\Gamma^a_{\hphantom{a}bc}\theta^c$ the set of connection one-forms. 
%obeying the requirement of vanishing torsion
%\begin{equation}
%\mathcal{ T}^a=\text{d}\theta^a + \omega^a_{\hphantom{a}b}\wedge \theta^b= \frac{1}{2}
 %   T^a_{\hphantom{a}bc} \theta^b\wedge \theta^c=0.\label{PMP-torsles}
%\end{equation}
%The cyclic and Bianchi identities ($\text{d}\wedge \text{d}  \theta^a=\text{d}\wedge \text{d}  \omega^a_{\hphantom{a}b} =0$), assuming a torsionless connection,  read:
%\begin{eqnarray}
%\mathcal{R}^a_{\hphantom{a}b} \wedge
 %   \theta^b&=&0,\\
%    \text{d}\mathcal{R}^a_{\hphantom{a}b}
%    + \omega^a_{\hphantom{a}c} \wedge \mathcal{R}^c_{\hphantom{c}b} - \mathcal{R}^a_{\hphantom{a}c} \wedge
%    \omega^c_{\hphantom{c}b} &=&0.
%\end{eqnarray}
We will assume the basis $\{\theta^a\}$ to be orthonormal with respect to the metric
\begin{equation}\label{PMP-met}
 \text{d}s^2 = \delta_{ab} \theta^a \theta^b  ,
\end{equation}
and the connection to be torsionless and  metric -- this latter statement is equivalent to
%\begin{equation}\label{PMP-metcon}
 $\omega_{ab}=-\omega_{ba}$,
%\end{equation}
%The latter together with (\ref{PMP-torsles}) determine the connection.
where the connection satisfies
\begin{equation}
%\mathcal{ T}^a=
\mathrm{d}\theta^a + \omega^a_{\hphantom{a}b}\wedge \theta^b= 
%\frac{1}{2}  T^a_{\hphantom{a}bc} \theta^b\wedge \theta^c
0  .
\label{PMP-torsles}
\end{equation}

The general holonomy group in four dimensions  is $SO(4)$, and \eqref{PMP-met} is invariant under local transfromations $\Lambda(x)$ such that
\begin{equation}
\nonumber
\theta^{a\prime}= 
\Lambda^{-1\, a}_{\hphantom{-1\, a}b}\theta^{b}  ,
\end{equation}
under which the curvature two-form transform as\footnote{Note the transformation of the connection: $  \omega^{a\prime}_{\hphantom{a}b}=\Lambda^{-1\, a}_{\hphantom{-1\, a}c}
\omega^{c}_{\hphantom{c}d}
\Lambda^d_{\hphantom{d}b}+ \Lambda^{-1\, a}_{\hphantom{-1\, a}c} \text{d}\Lambda^c_{\hphantom{c}b}$.}  
 \begin{equation}
 \nonumber
\mathcal{R}^{a\prime}_{\hphantom{a}b}=
\Lambda^{-1\, a}_{\hphantom{-1\, a}c}
\mathcal{R}^{c}_{\hphantom{c}d}
\Lambda^d_{\hphantom{d}b}  .
\end{equation}
Both $\omega_{ab}$ and $\mathcal{R}_{ab}$ are antisymmetric-matrix-valued forms, belonging to the representation $\mathbf{6}$ of $SO(4)$.

Four dimensions is a  special case as $SO(4)$ is factorized into $SO(3)\times SO(3)$. Both connection and curvature forms are therefore reduced with respect to each $SO(3)$ factor as
$\mathbf{3}\times \mathbf{1} +\mathbf{1}\times \mathbf{3}$, where $\mathbf{3}$ and $\mathbf{1}$ are respectively the vector and singlet representations. The connection and curvature decomposition leads to ($\lambda, \mu,\nu,\ldots = 1,2,3$ and $\epsilon_{123}=1$):
\begin{eqnarray}
\Sigma_\lambda =\frac{1}{2}\left(\omega_{0\lambda } + \frac{1}{2} \epsilon_{\lambda \mu \nu }\omega^{\mu \nu }\right)  ,&\quad&
A_\lambda =\frac{1}{2}\left(\omega_{0\lambda } - \frac{1}{2} \epsilon_{\lambda \mu \nu }\omega^{\mu \nu }\right)  ,\label{PMP-sdcon}\\ 
\mathcal{S}_\lambda =\frac{1}{2}\left(\mathcal{R}_{0\lambda } + \frac{1}{2} \epsilon_{\lambda \mu\nu}\mathcal{R}^{\mu\nu}\right)  ,&\quad&
\mathcal{A}_\lambda =\frac{1}{2}\left(\mathcal{R}_{0\lambda } - \frac{1}{2} \epsilon_{\lambda \mu\nu}\mathcal{R}^{\mu\nu}\right)  .
\end{eqnarray}
Using this decomposition, \eqref{PMP-curv} reads:
\begin{equation}
\label{PMP-curvcon}
 \mathcal{S}_\lambda= \text{d} \Sigma_\lambda -\epsilon_{\lambda\mu \nu } \Sigma^\mu \wedge  \Sigma^\nu   , \quad
  \mathcal{A}_\lambda= \text{d} A_\lambda +\epsilon_{\lambda\mu\nu } A^\mu \wedge A^\nu   .
 \end{equation}
   
    Usually $\mathcal{S}$ and $\mathcal{A}$ are referred to as self-dual and anti-self-dual components of the Riemann curvature. This follows from the definition of the dual forms (supported by the fully antisymmetric symbol\footnote{A remark is in order here for $D=7$ and $8$.
The octonionic structure constants  $\psi_{\alpha\beta\gamma}\ \alpha,\beta,\gamma\in\{1,\ldots,7\}$ and the dual $G_2$-invariant antisymmetric symbol $\psi^{\alpha\beta\gamma\delta}$ allow to define a duality relation in 7 and 8 dimensions with respect to an $SO(7)\supset G_2$, and an
$SO(8)\supset \text{Spin}_7$ respectively. Note, however, that neither $SO(7)$ nor $SO(8)$ is factorized, as opposed to $SO(4)$.} $\epsilon_{abcd}$)
\begin{equation}
\nonumber
%\tilde{\omega}^a_{\hphantom{a}b} &=& \frac{1}{2}
 %   \epsilon^{a\hphantom{bc}d}_{\hphantom{a}bc}\omega^c_{\hphantom{c}d}  ,\\ 
    \tilde{\mathcal{R}}^a_{\hphantom{a}b} = \frac{1}{2}
    \epsilon^{a\hphantom{bc}d}_{\hphantom{a}bc}\mathcal{R}^c_{\hphantom{c}d}  ,
\end{equation}
borrowed from Yang--Mills.
%\footnote{Note the action of the duality on the components, as 
%$ \tilde{\omega}^a_{\hphantom{a}b} =\tilde{\Gamma}^a_{\hphantom{a}bc}\theta^c,    \tilde{\mathcal{R}}^a_{\hphantom{a}b}=\frac{1}{2}
%    \tilde{R}^a_{\hphantom{a}bcd} \theta^c\wedge \theta^d$:
%$$
%   \begin{array}{rcl}
%    \tilde{\Gamma}^a_{\hphantom{a}bc}&=& \frac{1}{2}\epsilon^{a\hphantom{be}f}_{\hphantom{a}be}\Gamma^{e}_{\hphantom{e}fc}  ,\\
%    \tilde{R}^a_{\hphantom{a}bcd}&=& \frac{1}{2}\epsilon^{a\hphantom{be}f}_{\hphantom{a}be}R^{e}_{\hphantom{e}fcd}  ,
 %   \end{array}
%$$
%and similarly for the Weyl part or the Riemann.}
Under this involutive operation, $\mathcal{S}$ remains unaltered whereas $\mathcal{A}$ changes sign. Similar relations hold for the components $(\Sigma, A)$ of the connection.
        
Following the previous reduction pattern, the basis of 6 independent two-forms  can be decomposed in terms of two sets of singlets/vectors with respect to the two $SO(3)$ factors:
\begin{eqnarray}
\phi^\lambda &=&\theta^0\wedge\theta^\lambda  + \frac{1}{2} \epsilon^\lambda _{\hphantom{\lambda }\mu \nu }\theta^\mu \wedge\theta^\nu   , %\label{PMP-2fs}
\nonumber
\\ 
\chi^\lambda &=&\theta^0\wedge\theta^\lambda  - \frac{1}{2} \epsilon^\lambda _{\hphantom{\lambda }\mu \nu }\theta^\mu \wedge\theta^\nu   .
%\label{PMP-2fa}
\nonumber
\end{eqnarray}
In this basis, the 6 curvature two-forms $\mathcal{S}$ and $\mathcal{A}$ are decomposed as
\begin{equation}
\nonumber
      \begin{pmatrix}
   \mathcal{S}\\ 
   \mathcal{A}
  \end{pmatrix}=    
  \frac{r}{2}  
   \begin{pmatrix}
    \phi\\
   \chi
  \end{pmatrix}   ,   
    \end{equation}
where the $6\times 6$ matrix $r$ reads:    
\begin{equation}
\label{PMP-AHS}
r= 
\begin{pmatrix}
    A&C^+\\
   C^-&B
  \end{pmatrix}    =
   \begin{pmatrix}
    W^+&C^+\\
   C^-&W^-
  \end{pmatrix} +\frac{s}{6}\, \mathbf{I}_6  .
    \end{equation}

The 20 independent components of the Riemann tensor are stored inside the symmetric matrix  $r$ as follows:
\begin{itemize}
\item $s=\text{Tr}\, r =2 \text{Tr}\, A=2\text{Tr}\, B=\nicefrac{R}{2}$ is the scalar curvature.  
\item  The  9 components of the traceless part of the Ricci tensor $S_{ab}=R_{ab}-\frac{R}{4}g_{ab}$ ($R_{ab}=R^{c}_{\hphantom{a}acb}$) are given in $C^+=\left(C^-\right)^{\text{t}}$ as
\begin{equation}
\nonumber
S_{00}=\text{Tr}\, C^+, 
\quad 
S_{0\lambda }= 
%\frac{1}{2} \epsilon_\lambda ^{\hphantom{\lambda }\mu \nu }\left(C^-_{\mu \nu }-C^+_{\mu \nu }\right) 
\epsilon_\lambda ^{\hphantom{\lambda }\mu \nu }C^-_{\mu \nu },
\quad 
S_{\lambda \mu }= C^+_{\lambda \mu }+C^-_{\lambda \mu }-\text{Tr}\, C^+ \delta_{\lambda \mu }.
    \end{equation}
\item The 5 entries of the symmetric and traceless  $W^+$  are the components of the self-dual Weyl tensor, while $W^-$ provides the corresponding 5 anti-self-dual ones.
\end{itemize}
 In summary,    
\begin{eqnarray}
 \mathcal{S}_\lambda &=&\mathcal{W}_\lambda ^++\frac{1}{12}s\phi_\lambda +\frac{1}{2}C^+_{\lambda \mu }\chi^\mu ,\label{PMP-scurv}\\ 
   \mathcal{A}_\lambda &=&\mathcal{W}_\lambda ^-+\frac{1}{12}s\chi_\lambda +\frac{1}{2}C^-_{\lambda \mu }\phi^\mu ,\label{PMP-acurv} 
\end{eqnarray}
where 
\begin{equation}
\nonumber
\mathcal{W}_\lambda ^+=\frac{1}{2}W^+_{\lambda \mu }\phi^\mu , \quad
\mathcal{W}_\lambda ^-
=\frac{1}{2}W^-_{\lambda \mu }\chi^\mu 
 \end{equation}
are the self-dual and anti-self-dual Weyl two-forms respectively.
 
Given the above decomposition, the following nomenclature is used  (see e.g. \cite{PMP-Eguchi:1980jx} for details):
% \subsection{Back to the Riemann tensor}
   \begin{description}
\item[Einstein] $C^{\pm}=0$ ($\Leftrightarrow R_{ab}=\tfrac{R}{4}g_{ab}$)
\item[Ricci flat]  $C^{\pm}=0,\quad s=0$ 
\item[Self-dual] $ \mathcal{A}=0 \Leftrightarrow \{W^-=0, \ C^{\pm}=0, \ s=0\}$  
%{\color{fux} $\Rightarrow$ Ricci flat}
\item[Anti-self-dual]  $ \mathcal{S}=0 \Leftrightarrow \{W^+=0, \ C^{\pm}=0, \ s=0\}$ 
%{\color{fux} $\Rightarrow$ Ricci flat}
\item[Conformally self-dual] $W^-=0$
\item[Conformally anti-self-dual] $W^+=0$
\item[Conformally flat]  $W^{+}=W^-=0$
\end{description}
%Note that self-dual and anti-self-dual geometries are called half-flat in the mathematical literature, whereas self-dual and anti-self-dual is meant to be conformally self-dual and anti-self-dual. 
Quaternionic spaces are Einstein and conformally self-dual (or anti-self-dual). Conformal self-duality can also be combined with  K\"ahler structure. In either case, remarkable integrable structures emerge. 

Quaternionic conditions can be elegantly implemented by introducing the \emph{on-shell Weyl tensor}, defined as the antisymmetric-matrix-valued two-from:
\begin{equation}
\label{PMP-oswt}
 \hat{\mathcal{W}}^{ab}=\mathcal{R}^{ab}+k^2 \theta^a\wedge \theta^b  .
\end{equation}
Decomposing the latter \emph{\`a la} Atiyah--Hitchin--Singer, we obtain:
\begin{eqnarray}
 \hat{ \mathcal{W}}^+_\lambda&=&\mathcal{S}_\lambda+\frac{k^2}{2}\phi_\lambda=\mathcal{W}_\lambda^+ +\frac{1}{12}\left(s+6k^2\right)\phi_\lambda+\frac{1}{2}C^+_{\lambda\mu}\chi^\mu  , \label{PMP-oswts}\\ 
 \hat{ \mathcal{W}}^-_\lambda&=&\mathcal{A}_\lambda+\frac{k^2}{2}\chi_\lambda=\mathcal{W}_\lambda^- +\frac{1}{12}\left(s+6k^2\right)\chi_\lambda+\frac{1}{2}C^-_{\lambda\mu}\phi^\mu  . \label{PMP-oswta}
\end{eqnarray}
A quaternionic space is such that either $\hat{ \mathcal{W}}^+$ or $\hat{ \mathcal{W}}^-$ vanish.

\subsection{The filling-in problem}
\label{PMP-sec:fip}

A round three-sphere is a positive-curvature, maximally symmetric Einstein space with $SU(2)\times SU(2)$ isometry.
Its metric can be expressed using the Maurer--Cartan forms of $SU(2)$:
\begin{equation}\label{PMP-eq:rs}
\text{d}\Omega^2_3= \big(\sigma^1\big)^2+\big(\sigma^2\big)^2+ \big(\sigma^3\big)^2
 \end{equation}
with 
\begin{equation}
\nonumber
  \begin{cases}
\sigma^1= \sin\vartheta \sin\psi \, \text{d}\varphi+\cos \psi \, \text{d}\vartheta \\
\sigma^2= \sin\vartheta\cos\psi\, \text{d}\varphi-\sin\psi\, \text{d}\vartheta\\
\sigma^3=\cos\vartheta\, \text{d}\varphi+\text{d}\psi;
\end{cases}
 \end{equation}
$0\leq\vartheta\leq \pi, 0\leq\varphi\leq 2\pi, 0\leq\psi\leq {4\pi}$ are the Euler angles.

A hyperbolic four-space $H_4$ is a negative-curvature, maximally symmetric Einstein space. It is a foliation over three-spheres and its metric reads:
\begin{equation}
\nonumber
\text{d}s^2_{H_4}=\frac{\text{d}r^2}{1+k^2r^2} + k^2r^2 \text{d}\Omega^2_3.
\end{equation}
(we assumed $R_{ab}=-3k^2 g_{ab}$ for $H_4$).
The conformal boundary of $H_4$ is reached at $r\to \infty$ as
\begin{equation}
\nonumber
\text{d}s^2_{H_4} \underset{r\to \infty}{\longrightarrow} k^2r^2 \text{d}\Omega^2_3  .
\end{equation}
In this sense, the round three-sphere  is filled-in with $H_4$, the latter being the only regular metric filling-in this three-dimensional space. 

The natural question to ask in view of the above is how to fill-in the more general Berger sphere $S^3$,  which is  a homogeneous but non-isotropic deformation of \eqref{PMP-eq:rs}:
\begin{equation}\label{PMP-eq:BS}
\text{d}\Omega^2_{S^3}= \big(\sigma^1\big)^2+\big(\sigma^2\big)^2+4n^2 k^2\big(\sigma^3\big)^2
 \end{equation}
with $nk$ constant. This metric is invariant under $SU(2) \times U(1)$, respectively generated by the Killings
\begin{equation}
  \begin{cases}
\nonumber
%  \label{PMP-LKil}
\xi_1= - \sin\varphi \cot\vartheta\, \partial_\varphi+\cos \varphi\, \partial_\vartheta+\frac{\sin \varphi}{\sin \vartheta} \, \partial_\psi \\
\xi_2=  \cos\varphi \cot\vartheta\, \partial_\varphi+\sin \varphi \,\partial_\vartheta-\frac{\cos \varphi}{\sin \vartheta} \, \partial_\psi \\
\xi_3= \partial_\varphi,
\end{cases}
\end{equation}
and $\partial_\psi$.

LeBrun studied the filling-in problem
in general terms \cite{PMP-LeBrun82} and showed that an analytic three-metric can be regularly filled-in by a four-dimensional Einstein space that has self-dual (or anti-self-dual) Weyl tensor, \emph{i.e.} by a \emph{quaternionic} space. In modern holographic words, LeBrun's result states that requiring regularity makes the boundary metric a sufficient piece of data for reconstructing the bulk. Regularity translates into conformal self-duality, which effectively reduces by half the independent Cauchy data of the problem, as we will see in Sect. \ref{PMP-sec:wsd}.

\subsection{A concrete example}\label{PMP-sec:ce}

LeBrun's analysis is very general. We can illustrate it in the specific example of the Berger sphere $S^3$. %given in \eqref{PMP-eq:BS}.  
We search therefore a four-dimensional foliation over $S^3$, which is Einstein. This leads to the Bianchi IX Euclidean Schwarzschild--Taub--NUT family on hyperbolic space (\emph{i.e.} with $\Lambda=-3k^2 $):
\begin{equation}\label{PMP-hyptaubnut}
\text{d}s^2 =
 \frac{\text{d}r^2}{V(r)}
+\left(r^2-n^2\right)\left(\big(\sigma^1\big)^2+
\left(\sigma^2\right)^2\right)+4n^2V(r)\left(\sigma^3\right)^2
 \end{equation}
with
\begin{equation}
\label{PMP-hyptaubnutpot}
V(r)= \frac{1}{r^2-n^2}\left[
r^2+n^2 -2 M r +k^2\left(
r^4-6n^2r^2 -3 n^4
\right)
\right]  ,
\end{equation}
where $M$ and $n$ are the mass and nut charge. Clearly the metric fulfills the boundary requirement since 
\begin{equation}
\nonumber
\text{d}s^2 \underset{r\to \infty}{\longrightarrow} r^2 \text{d}\Omega^2_{S^3}  ,
\end{equation}
where $\text{d}\Omega^2_{S^3}$ is given in \eqref{PMP-eq:BS}.

The family of solutions at hand depends on 2 parameters, $M$ and $n$, of which only the second remains visible on the conformal boundary. In that sense, the bulk is not fully determined by the boundary metric.
However, regularity is not always guaranteed either, as $\text{d}s^2$ is potentially singular at $r=+ n$ or $r=- n$ (depending on whether the range for $r$ is chosen positive or negative). 
Actually, this locus coincides with the fixed points of the Killing vector $\partial_{\psi}$, generating the extra $U(1)$.\footnote{In four dimensions, the fixed locus of an isometry is either a zero-dimensional or a two-dimensional space. The first case corresponds to a nut, the second to a bolt, and both can be removable singularities under appropriate conditions (see \cite{PMP-nuts} for a complete presentation).} In the present case, these are nuts and they are removable provided the space surrounding them is locally flat. 

In order to make the above argument clear, let us focus for concreteness on $r=n$ (assuming thus $r>0$), write $r=n+\epsilon$ and expand the metric using momentarily $\epsilon$ as radial coordinate:
\begin{eqnarray}
\text{d}s^2 &\approx& \frac{\text{d}\epsilon^2}{V(n)+\epsilon V'(n)}
+2n\epsilon
\left(
 \text{d}\vartheta^2 +
\sin^2\vartheta \text{d}\varphi^2
\right)\nonumber\\
\label{PMP-hyptaubnut-exp}
&&+4n^2\left(V(n)+\epsilon V'(n)\right)\left( \text{d}\psi + \cos \vartheta  \text{d}\varphi\right)^2  .
\end{eqnarray}
Clearly to reconstruct locally flat space we must impose $V(n)=0$ and $V'(n)=\nicefrac{1}{2n}$. The first of these requirements is equivalent to 
\begin{equation}
\label{PMP-sd-QTN}
M= n\left(1-4k^2n^2\right)  ,
\end{equation}
and makes  the second automatically satisfied. Under \eqref{PMP-sd-QTN} and with $\tau=2\sqrt{2n\epsilon}$ (proper time), Eq. \eqref{PMP-hyptaubnut-exp}  reads: 
\begin{equation}
\nonumber
%\label{PMP-hyptaubnut-exp-cond}
\text{d}s^2 \approx \text{d}\tau^2 + \frac{\tau^2}{4}\left(\text{d}\psi^2+\text{d}\varphi^2+ \text{d}\vartheta^2 + 2 \cos \vartheta\,  \text{d}\psi\,  \text{d}\varphi \right) ,
\end{equation}
which is indeed $\mathbb{R}^4$. 

We can similarly analyze the behavior around $r=-n$.  We then reach the same conclusion, with an overall change of sign in condition \eqref{PMP-sd-QTN}. These conditions are nothing but conformal (anti-)self-duality requirements, as we see by computing the Weyl components of the curvature, $W^\pm$, in the decomposition \eqref{PMP-AHS}:
\begin{equation}
\nonumber
W^\pm=\frac{M\mp n(1-4k^2n^2)}{(r\mp n)^3}\begin{pmatrix}-1&0&0\\
0&-1&0\\
0&0&2
\end{pmatrix}  .
\end{equation}
The regularity requirement for the family of Einstein spaces \eqref{PMP-hyptaubnut} is thus equivalent to demand the space be quaternionic. In that case, the boundary metric contains enough information for determining the bulk and solving thereby the filling-in problem for the Berger sphere. 

For the quaternionic Schwarzschild--Taub--NUT geometries \eqref{PMP-hyptaubnut} with \eqref{PMP-sd-QTN}, the function $V(r)$ in \eqref{PMP-hyptaubnutpot} reads:
\begin{equation}
\nonumber
%\label{PMP-hyptaubnutpot-sd}
V(r)= \frac{r-n}{r+n}\left[1+k^2
(r-n)(r+3n)
\right].
\end{equation}
These geometries belong to the general class of Calderbank--Pedersen \cite{PMP-CP02}, which is the family of quaternionic spaces with at least two commuting Killing fields.\footnote{The metrics at hand are sometimes called \emph{spherical Calderbank--Pedersen}, because they possess in total four Killings, of which three form an $SU(2)$ algebra.}  They belong to a wide web of structures, and are in particular conformal to a family of spaces, which are K\"ahler and Weyl-anti-self-dual with vanishing scalar curvature, known as LeBrun geometries \cite{PMP-LeBrun-88}. The limit $n\to \infty$ deserves a particular attention, as it corresponds to the pseudo-Fubini--Study\footnote{This is the non-compact Fubini--Study. The ordinary Fubini--Study corresponds to the compact $\mathbb{CP}_2=\frac{SU(3)}{U(2)}$ and has positive cosmological constant.} metric on $\widetilde{\mathbb{CP}_2}=\frac{SU(2,1)}{U(2)}$. Further holographic properties of these geometries can be found in \cite{PMP-Zoubos:2002cw, PMP-Zoubos:2004qm}.

\section{Weyl self-duality from the boundary}\label{PMP-sec:sdb}

The filling-in problem was presented as the ancestor of holography in the sense that (\romannumeral1) it poses the problem of reconstructing the bulk out of the boundary and (\romannumeral2) it raises the issue of regularity as a mean to relate \emph{a priori} independent boundary data. The bonus is that in the present Euclidean approach, regularity condition appears as conformal self-duality requirement, which in turn makes Einstein's equations integrable and the bulk an exact solution.

The natural question to ask at this stage is how the bulk Weyl self-duality gets manifest on the boundary. In order to answer, we must perform a clear analysis of the independent boundary data following Fefferman--Graham approach and recast in these data the self-duality requirement.

\subsection{The Fefferman--Graham expansion}

The work of LeBrun \cite{PMP-LeBrun82}, quoted previously in the framework of the filling-in problem, led Fefferman and Graham to set up a systematic expansion for Einstein metrics in powers of a radial coordinate \cite{PMP-FG1, PMP-FG2}. The infinite set of coefficients are data of the boundary, expressed in terms of two independent ones: $g_{\mu\nu}$ and $F_{\mu\nu}$. From a Hamiltonian perspective, with the radial coordinate as evolution parameter, $g_{\mu\nu}$ and $F_{\mu\nu}$ are Cauchy data of ``coordinate'' and ``momentum'' type. The former is of geometric nature, the latter is not. In the holographic language, $g_{\mu\nu}$ corresponds to a non-normalizable mode and is the boundary metric, whereas 
$F_{\mu\nu}$ is related to a normalizable operator and carries information on the energy--momentum-tensor expectation value of the boundary field theory:
\begin{equation}
\label{PMP-FT}
T_{\mu\nu}=\frac{3k}{8\pi G_{\text{N}}}F_{\mu\nu}  ,
\end{equation}
where $G_{\text{N}}$ is four-dimensional Newton's constant. 
%($k$ is related to the cosmological constant: $\Lambda = -3 k^2$).

The method of Fefferman--Graham is well suited for holography and has led to important developments (see e.g. \cite{PMP-SS, PMP-dHSS, PMP-Papadimitriou:2005ii}). It nicely fits the gravito-electric/gravito-magnetic split Hamiltonian formalism of four-dimensional gravity \cite{PMP-Mansi:2008br, PMP-Mansi:2008bs}. In the Euclidean, this formalism is basically adapted to the self-dual/anti-self-dual splitting of the gravitational degrees of freedom presented in Sect. \ref{PMP-sec:CDD}.

Let us summarize here the basic facts, leaving aside the rigorous and complete exhibition that can be found in the above references. In Palatini formulation, the four-dimensional (bulk) Einstein--Hilbert action reads:
\begin{equation}
\nonumber
  I_{\text{EH}} 
  = -\frac{1}{32\pi G_{\text{N}}}
   \int_{\mathcal{M}}
   \epsilon_{abcd}\left(
    \mathcal{R}^{ab}+\frac{k^2}{2}  \theta^a\wedge \theta^b\right) \wedge \theta^c\wedge \theta^d.
\end{equation}
As we already mentioned, $\theta^a$, $a=r,\lambda$ are basis elements of a coframe, orthonormal with respect to the signature $(+\eta++)$. The first direction $r$ is the holographic one, and $\text{x}\equiv(t, x^1,x^2)$ are the remaining coordinates, surviving on the conformal boundary -- with $t\equiv x^3$ in the Euclidean instance ($\eta=+$). 

The most general form for the coframe is 
\begin{equation}
\nonumber
 \theta^r=N\frac{\text{d}r}{kr}  ,\quad 
 \theta^\lambda=N^\lambda\text{d}r+\tilde\theta^\lambda  ,
\end{equation}
whereas the Levi--Civita connection generally reads: 
\begin{equation}
\nonumber
\omega^{r\lambda}=q^{r\lambda}\text{d}r+\mathcal{K}^\lambda  ,\quad 
\omega^{\mu\nu}=- \epsilon^{\mu\nu\lambda}\left(Q_\lambda\frac{\text{d}r}{kr}+\mathcal{B}_\lambda\right)  .
\end{equation}
Without loss of generality, we can make the following gauge choice:
\begin{equation}
\nonumber
N=1 , \quad N^\mu= q^{r\mu}=Q_\rho=0  ,
\end{equation}
leading to the Fefferman--Graham form for the bulk metric:
\begin{equation}
\label{PMP-FGf}
\text{d}s^2=\frac{\text{d}r^2}{k^2r^2}+\eta_{\mu\nu} \tilde\theta^\mu \tilde\theta^\nu  .
\end{equation}
The connection is encapsulated in $\mathcal{K}^\mu$ and $\mathcal{B}_\lambda$. In Euclidean signature ($\eta =+$), these are vector-valued (with respect to the holonomy $SO(3)$ subgroups) connection one-forms, related  to the (anti-)self-dual ones introduced in \eqref{PMP-sdcon}:
\begin{equation}\label{PMP-eq:KB}
\mathcal{K}_\lambda=A_\lambda+\Sigma_\lambda  , \quad 
\mathcal{B}_\lambda=A_\lambda-\Sigma_\lambda  .
\end{equation}
The zero-torsion condition \eqref{PMP-torsles} translates in this language into 
\begin{equation}
\label{PMP-ztc-asd}
  \begin{cases}
\mathcal{K}_\lambda\wedge \tilde\theta^\lambda=0  \\
\text{d}\tilde\theta^\lambda=\frac{1}{kr} \mathcal{K}^\lambda\wedge \text{d}r-\epsilon^{\lambda\mu\nu}
\mathcal{B}_\mu\wedge  \tilde\theta_\nu  .
\end{cases}
\end{equation}

With the present choice of gauge, all relevant information on the bulk geometry is stored inside $\left\{\tilde\theta^\lambda,\mathcal{K}^\lambda, \mathcal{B}^\lambda \right\}$. Assuming the metric be Einstein, leads to a very specific $r$-expansion of these vector-valued one-forms, in terms of the boundary data. This is the Fefferman--Graham expansion:
\begin{eqnarray}
\tilde\theta^\lambda(r,\text{x})&=& kr\, E^\lambda(\text{x})+\sum_{\ell=0}^\infty\frac{1}{(kr)^{\ell+1}}F^\lambda_{[\ell+2]}(\text{x})  ,
%+\frac{1}{k^2r^2}F^\lambda(\text{x}) +\frac{1}{kr} F^\lambda_{[2]}(\text{x})+\frac{1}{k^2r^2}F^\lambda(\text{x})+\cdots 
\label{PMP-bulkfr}\\
\mathcal{K}^\lambda(r,\text{x}) &=& -k^2 r\, E^\lambda(\text{x})+k\sum_{\ell=0}^\infty\frac{\ell+1}{(kr)^{\ell+1}}
F^\lambda_{[\ell+2]}(\text{x})  ,
%+\frac{2}{kr^2}F^\lambda(\text{x})+\cdots
\label{PMP-bulkel} \\
\mathcal{B}^\lambda(r,\text{x}) &=& B^\lambda(\text{x})+\sum_{\ell=0}^\infty\frac{1}{(kr)^{\ell+2}}B^\lambda_{[\ell+2]}(\text{x})  .
%+\cdots 
\label{PMP-bulkmag}
\end{eqnarray}
The boundary data are vector-valued one-forms. They are not all independent, and higher orders are derivatives of lower orders (we will meet an example of this ``horizontal'' relationship in a short while). Furthermore, due to the zero-torsion condition \eqref{PMP-ztc-asd}, further ``vertical'' relations exist order by order amongst the three sets. This is manifest when comparing \eqref{PMP-bulkfr} and \eqref{PMP-bulkel}, where the relations are algebraic. The forms in \eqref{PMP-bulkfr} and \eqref{PMP-bulkmag} are also related, in a differential manner, though. 

The form $E^\lambda$ is the boundary coframe. It is the first independent coefficient and it allows to reconstruct the three-dimensional boundary metric:
\begin{equation}
\nonumber
 \text{d}s^2_{\text{bry.}}=\lim_{r\to \infty}\frac{\text{d}s^2}{k^2r^2}= \eta_{\mu\nu}E^\mu E^\nu.
\end{equation}
The one-form $B^\mu$ appearing in the expansion of the magnetic component of the bulk connection, Eq. \eqref{PMP-bulkmag},  is the boundary Levi--Civita connection, differentially related to the coframe (boundary zero-torsion condition):
\begin{equation}
\nonumber
\text{d}E^\lambda=\epsilon^{\lambda\mu\nu} B_\nu\wedge E_\mu.
\end{equation}
Other forms such as $F^\mu_{[2]}=F^\mu_{[2]\nu}E^{\nu}$ or $B^\mu_{[2]}=B^\mu_{[2]\nu}E^{\nu}$ are also geometric, respectively related to the Schouten and Cotton--York tensors:\footnote{In three dimensions, the Schouten tensor is defined as $S^{\mu\nu}=R^{\mu\nu}-\frac{R}{4}g^{\mu\nu}$, whereas the Cotton--York tensor is the Hodge-dual of the Cotton tensor, defined in Eq. \eqref{PMP-cotdef}.
%$C^{\mu\nu}=\eta^{\mu\rho\sigma} \nabla_\rho\left(R^{\nu}_{\hphantom{\nu}\sigma}-\frac{1}{4}R\delta^{\nu}_\sigma \right)$. 
The latter replaces the always vanishing three-dimensional Weyl tensor. In particular, conformally flat boundaries have zero Cotton tensor and vice versa.}
\begin{equation}
\label{PMP-F2B2}
S^{\mu\nu}=-2k^2F_{[2]}^{\mu\nu}  ,\quad 
C^{\mu\nu}=2k^2B_{[2]}^{\mu\nu}  .
\end{equation}
There is again a differential relationship among the two, following basically from the bulk zero-torsion condition \eqref{PMP-ztc-asd}, since by definition
\begin{equation}
\label{PMP-cotdef}
C^{\mu\nu}=\eta^{\mu\rho\sigma}
\nabla_\rho S^{\nu}_{\hphantom{\nu}\sigma} 
\end{equation}
($\eta^{\mu\rho\sigma}=\nicefrac{\epsilon^{\mu\rho\sigma}}{\sqrt{\vert g \vert}}$).

Other curvature tensors of arbitrary order appear in the Fefferman--Graham expansion, all differentially related to the ones already described above. These tensors do \emph{not} exhaust, however, all coefficients of the series \eqref{PMP-bulkfr}, \eqref{PMP-bulkel} and \eqref{PMP-bulkmag}, as some infinite sequences of those are not of geometric nature, \emph{i.e.} are not determined by the boundary metric itself (or by the coframe $E^\mu$). Instead, they follow differentially from the second independent piece of data, $F^\mu\equiv F^\mu_{[3]}$, related to the energy--momentum expectation value according to \eqref{PMP-FT}.  The interested reader will find a more complete exhibition of the Fefferman--Graham expansion in the literature, and particularly in \cite{PMP-Mansi:2008br, PMP-Mansi:2008bs} for the gravito-electric/gravito-magnetic split formalism.

\subsection{Self-duality and its Lorentzian extension}
\label{PMP-sec:wsd}

 \paragraph{Riemann self-duality}

A word on Riemann self-duality is in order at this stage, before exploring the more subtle issue of Weyl self-duality.

Demanding the Riemann tensor be (anti-)self-dual (see end of Sect. \ref{PMP-sec:CDD}) guarantees Ricci flatness and Weyl (anti-)self-duality. Such a requirement on the curvature is easily transported to the connection, using Eq. \eqref{PMP-curvcon}: the anti-self-dual connection $\mathcal{K}_\mu+\mathcal{B}_\mu$
(see Eq. \eqref{PMP-eq:KB})
of a self-dual Riemann is either vanishing or a pure gauge (flat).  This basically removes the corresponding degrees of freedom and gives an easy way to handle the problem via first-order differential equations. 

The case of Bianchi foliations along the radial (holographic) direction, as the example we described in Sect. \ref{PMP-sec:ce}, has been largely analyzed in the literature (see \cite{PMP-Eguchi:1980jx} for a general discussion, \cite{PMP-Gibbons:1979xn} for Bianchi IX, or \cite{PMP-Bourliot:2009fr, PMP-Bourliot:2009ad, PMP-Petropoulos:2011qq} for a more recent general and exhaustive Bianchi analysis). The requirement of (anti-)self-dual Riemann leads to the following equation:
 \begin{equation}\label{PMP-flsdLC}
\mathcal{K}_\mu\pm\mathcal{B}_\mu
=\lambda_{\mu\nu}\sigma^\nu  ,
\end{equation}
where $\sigma^\nu$ are the Maurer--Cartan forms of the Bianchi group, and $\lambda_{\mu\nu}$ a constant matrix parameterizing the homomorphisms mapping $SO(3)$ onto the Bianchi group. Expressing  $\mathcal{K}_\mu$ and $\mathcal{B}_\mu$ in terms of the metric, \eqref{PMP-flsdLC} provides a set of first-order differential equations that have usually remarkable integrability properties. For concreteness, in the case of Bianchi IX ($SO(3)$) foliations,  $\lambda_{\mu\nu} =0$ or $\delta_{\mu\nu}$. The former case leads to the Lagrange equations, whereas the latter to the Darboux--Halphen system. Both systems are integrable, with celebrated solutions such as Eguchi--Hanson or BGPP for the first \cite{PMP-Eguchi:1978gw, PMP-Eguchi:1979yx, PMP-Belinsky:1978ue}, and Taub--NUT or Atiyah--Hitchin for the second \cite{PMP-Taub-nut, PMP-Atiyah1}.

 \paragraph{Weyl self-duality}

Demanding Weyl (anti-)self-duality  is not sufficient for setting $\mathcal{K}_\mu\pm\mathcal{B}_\mu$ 
as a pure gauge (flat connection). In the case of Bianchi foliations e.g. Eq. \eqref{PMP-flsdLC} is still valid but 
$\lambda_{\mu\nu}$ is a function of the radial coordinate $r$, and satisfies a first-order differential equation. The general structure of this equation (independently of any ansatz such as a Bianchi foliation) imposes a certain behavior and this is how Weyl (anti-)self-duality affects boundary conditions in a way that becomes transparent in the Fefferman--Graham large-$r$ expansion.

We are specifically interested in quaternionic spaces, which are Einstein and conformally (anti-)self-dual. 
Thanks to the on-shell Weyl tensor \eqref{PMP-oswt},  these requirements 
are simply either  $\hat{ \mathcal{W}}^+_\lambda=0$ (anti-self-dual) or  $\hat{ \mathcal{W}}^-_\lambda=0$ (self-dual). Expressions \eqref{PMP-oswts} and \eqref{PMP-oswta}, combined with \eqref{PMP-curvcon} and \eqref{PMP-eq:KB}--\eqref{PMP-bulkmag}, allow to establish the effect of Weyl (anti-)self-duality on the boundary one-forms. This appears as a hierarchy of algebraic equations\footnote{When dealing with the Fefferman--Graham expansion together with Einstein dynamics, attention should be payed to the underlying variational principle. This sometimes requires Gibbons--Hawking boundary terms to be well posed. In the Hamiltonian language, these terms are generators of canonical transformations and in AdS/CFT their effect is known as holographic renormalization. These subtleties are discussed in  \cite{PMP-Mansi:2008br, PMP-Mansi:2008bs, PMP-Leigh:2007wf, PMP-deHaro:2007fg, PMP-deHaro:2008gp, PMP-Miskovic:2009bm}, together with the specific role of the Chern--Simons boundary term, which produces the boundary Cotton tensor, and in conjunction with Dirichlet vs. Neumann boundary conditions. One should also quote the related works \cite{PMP-Bakas:2008gz, PMP-Bakas:2008zg}, in the linearized version of gravitational duality though.}  
\begin{equation}
\nonumber
%\label{PMP-sdl}
k\left[(\ell+2)^2-1\right] F^\lambda_{[\ell+3]}\pm(\ell+2)B^\lambda_{[\ell+2]}=0  , \quad \forall \ell \geq 0
\end{equation}
(the upper $+$ sign corresponds to the self-dual case), of which only the first is independent:
\begin{equation}\label{PMP-sd0}
3k F^\lambda_{[3]}\pm 2 B^\lambda_{[2]}=0 .
\end{equation}
The others follow from the already existing horizontal differential relationships. This \emph{algebraic} equation between \emph{a priori} independent boundary data is at the heart of conformal self-duality. In terms of the boundary energy--momentum and Cotton tensors (see \eqref{PMP-FT} and \eqref{PMP-F2B2}), Eq.  \eqref{PMP-sd0} reads:
\begin{equation} \label{PMP-sdb} 
8\pi G_{\text{N}}k^2 T^{\mu\nu} \pm C^{\mu\nu} =0
  .
\end{equation}

Several important comments are in order at this stage. Firstly, referring to the original problem of Sect. \ref{PMP-sec:fip}, Eq. \eqref{PMP-sd0} provides the filling-in boundary condition for some \emph{a priori} given boundary metric (not necessarily a three-sphere as originally studied in \cite{PMP-LeBrun82}).  This condition tunes algebraically the Cauchy data (``initial position'' and ``initial momentum''), in such a way that any boundary metric can be filled-in \emph{regularly}. Following the intuition developed in the example of Sect. \ref{PMP-sec:ce}, we may slightly relax this condition and trade it for 
\begin{equation}\label{PMP-sdw}
wT^{\mu\nu} + C^{\mu\nu} =0 ,
\end{equation}
where we now allow for any real $w$ and not solely $w=\pm 8\pi G_{\text{N}} k^2$. The filling-in is still expected to occur, without guaranty for the regularity though. 

Secondly, as discussed in the introduction, duality is underlying integrability. This statement is clear in the case of Riemann self-duality, where the key is the reduction of the differential order of the equations. For conformal self-duality it operates via an appropriate tuning of the boundary conditions, the effect of which would be better qualified as exactness rather than integrability: the equations of motion are not simplified, but the initial conditions select a specific corner of the phase space, which enables for exact solutions to emerge, \emph{i.e.} for the Fefferman--Graham series to be resummable. Furthermore, even though self-duality (Riemann or Weyl) does not apply to the Lorentzian frame,\footnote{In four-dimensional metrics with Lorentzian signature, self-duality leads either to complex solutions, or to Minkowski and $\text{AdS}_4$, which are both self-dual \emph{and} anti-self-dual (they have vanishing Riemann and vanishing Weyl, respectively).} condition \eqref{PMP-sdw} remains consistent for a Lorentzian boundary, and is expected, following our heuristic arguments, to guarantee the resummability of the Fefferman--Graham expansion and lead to exact solutions. This is not a theorem, much like everything regarding the relationship between integrability and self-duality in general, but the idea seems to work, as we will see in Sect. \ref{PMP-pCg}.

Our last comment concerns the potential developments around Eq. \eqref{PMP-sdw}, already announced, and discussed, in the introduction (Eq. \eqref{PMP-eq:sdc}). This equation is a boundary condition, which, however, can follow from a three-dimensional variational principle. In order to enforce it via this principle, we must equip the boundary field theory with specific dynamics that incorporates three-dimensional gravity, in the form of a topological massive term, as suggested by Eqs. \eqref{PMP-eq:sdcd} and \eqref{PMP-eq:sdca}. The first term in $S$ is the phenomenological holographic matter action, whereas the second is the Chern--Simons term with $\omega_3$ the Lagrangian density given in terms of the boundary connection one-form $\gamma$:
\begin{equation}
\nonumber
\omega_3(\gamma)= {1 \over 2} \mathrm{Tr}\left(\gamma\wedge \mathrm{d}\gamma+
\frac{2}{3}\gamma\wedge\gamma\wedge\gamma\right)   .
\end{equation}
Conceptually, this is a non-trivial step as holography is not supposed \emph{a priori} to endow the boundary theory with gravitational dynamics. It raises three questions:
\begin{enumerate}
\item What are the allowed boundary geometries, given certain assumptions on the energy--momentum tensor ?
\item What are the bulk geometries that reproduce holographically the boundary data? Are those exact Einstein spaces, \emph{i.e.} is the corresponding Fefferman--Graham expansion resummable in accordance with the above discussion ?
\item Are there situations where gravitational degrees of freedom emerge? 
\end{enumerate}
We will answer questions 1 and 2, at least in some specific framework, leaving open interesting extensions. As we will see, in some situations, the boundary geometry is really a topologically massive gravity vacuum -- as if the three-dimensional Einstein--Hilbert term were effectively present in \eqref{PMP-eq:sdca}. We will not delve into question 3, because this is a definitely different direction of investigation. The interested reader may find Ref. \cite{PMP-deHaro:2008gp} useful and inspiring regarding that issue.

\section{Application to holographic fluids}

The purpose of the present part is to answer questions 1 and 2 raised in Sect. \ref{PMP-sec:wsd}. Solving Eq. \eqref{PMP-sdw} is possible, provided some assumptions are made both on the energy--momentum tensor, and on the boundary metric. These assumptions are motivated by our goal to probe transport coefficients for holographic fluids, without performing linear-response analysis. For that we must study equilibrium configurations of the fluid in various exact non-trivial backgrounds and design accordingly the boundary data. These satisfy Eq. \eqref{PMP-sdw} and are integrable \emph{i.e.} the corresponding Fefferman--Graham expansion is resummable. 

\subsection{Fluids at equilibrium in Papapetrou--Randers backgrounds}
\label{PMP-FEPR}

 \paragraph{Hydrodynamic description} 

A given bulk configuration (geometry possibly supplemented with other fields) provides a boundary geometry, and a finite-temperature and finite-density state of the -- generally unknown -- microscopic boundary theory. It has expectation value $T_{\mu\nu}$ for the energy--momentum tensor, satisfying 
\begin{equation}\label{PMP-cons}
\nabla_\mu T^{\mu\nu}=0 , 
\end{equation}
and possibly other conserved currents. This state may be close to a hydrodynamic configuration and is potentially described within the hydrodynamic approximation. This assumes, among others,  local thermodynamic equilibrium. For this description to hold, it is necessary that the scale of variation of the diverse quantities describing the fluid be large compared to any microscopic scale (such as the mean free path).  We will work in this framework and furthermore suppose the fluid neutral, as the only bulk degrees of freedom are gravitational in our case. 

The relativistic fluid is described in terms of a velocity field $\text{u}(\text{x})$, as well as of
local thermodynamic quantities like $T(\text{x}), p(\text{x}), \varepsilon(\text{x}), s(\text{x})$, obeying an equation of state and thermodynamic identities 
\begin{equation}
\nonumber
%\label{PMP-thermo}
\begin{cases}
sT=\varepsilon+ p\\ 
\text{d}\varepsilon=T\text{d}s.
\end{cases}
\end{equation}
All these enter the energy--momentum tensor.
 The energy--momentum tensor of a neutral hydrodynamic system can be expanded in derivatives of the hydrodynamic variables, namely
\begin{equation}
\label{PMP-T00}
T^{\mu\nu}=T^{\mu\nu}_{(0)} +T^{\mu\nu}_{(1)}+T^{\mu\nu}_{(2)}+\cdots,
\end{equation}
where the subscript denotes the number of covariant derivatives. The validity of this derivative expansion is subject to the above assumptions regarding the scale of variation.
The zeroth order energy--momentum tensor is the so called perfect-fluid energy--momentum tensor:  
\begin{equation}\label{PMP-T0}
T_{(0)}^{\mu\nu}=  \varepsilon u^\mu u^\nu +p\,
\Delta^{\mu\nu},
\end{equation}
where $\Delta^{\mu\nu} = u^\mu u^\nu + g^{\mu\nu}$ is the projector onto the space orthogonal to $\mathrm{u}$. This corresponds to a fluid being locally in equilibrium, in its proper frame.\footnote{Defining the local proper frame, \emph{i.e.} the velocity field $\text{u}$, is somewhat ambiguous in relativistic fluids. A  possible choice is the \emph{Landau frame}, where the non-transverse part of the energy--momentum tensor vanishes when the pressure is zero. This will be our choice.} 
The conservation of the perfect-fluid energy--momentum tensor leads to the relativistic Euler equations:
\begin{equation}\label{PMP-Euler0}
 \begin{cases}
 \nabla_{\mathrm{u}}\varepsilon+ (\varepsilon+p)\Theta=0
 \\\nabla_{\perp}p +(\varepsilon+p)\mathrm{a}=0,
\end{cases}
\end{equation}
where $\nabla_{\mathrm{u}}= \mathrm{u}\cdot\nabla$, $\Theta  = \nabla\cdot \mathrm{u}$, $\nabla_{\perp\mu} = \Delta_\mu^{\phantom{\mu}\nu}\nabla_\nu$, and $\text{a} = \mathrm{u}\cdot\nabla \text{u}$
(more formulas on kinematics of relativistic fluids are collected in App. \ref{PMP-vfc}).

The higher-order corrections to the energy--momentum tensor involve the transport coefficients of the fluid. These are phenomenological parameters that encode the microscopic properties of the underlying system. Listing them order by order requires to classify all transverse tensors (possibly limited to traceless and Weyl-covariant if the microscopic theory is conformally invariant) and this depends on the space--time dimension.\footnote{We recommend Refs. \cite{PMP-Romatschke:2009im, PMP-Kovtun:2012rj} for a recent account of that subject. Insightful information was also made available  thanks to the developments on fluid/gravity correspondence \cite{PMP-Hubeny, PMP-Rangamani:2009xk}.}
In the context of field theories, the transport coefficients can be determined from studying correlation functions of the energy--momentum tensor at finite temperature in the low-frequency and low-momentum regime (see for example \cite{PMP-Moore:2010bu}). 

 \paragraph{Equilibrium and perfect equilibrium} 

Studying fluids at \emph{equilibrium} on \emph{non-trivial} backgrounds can provide information on their transport properties. A fluid in global thermodynamic equilibrium\footnote{This should not be confused with a steady state, where we have stationarity due to a balance between external driving forces and internal dissipation. Such situations will not be discussed here.} is described by a 
stationary solution\footnote{It is admitted that a non-relativistic fluid is stationary when its velocity field is time-independent. This is of course an observer-dependent statement. For relativistic fluids, one could make this more intrinsic saying that the velocity field commutes with a globally defined time-like Killing vector, assuming that the later exists. Note also that statements about global thermodynamic equilibrium in gravitational fields are subtle and the subject still attracts interest \cite{PMP-Banerjee:2012iz}.} of the relativistic equations of motion \eqref{PMP-cons}, assuming that such solutions exist. Finding solutions to these equations is generally a hard task, in particular because most of the transport coefficients are unknown. As it will become clear in a short while, the concept of \emph{perfect equilibrium} provides a natural way out, giving access to non-trivial information about transport properties.

The prototype example,
where global thermodynamic description applies,
is the one of an inertial fluid in Minkowski background with globally defined constant temperature, energy density and pressure. 
In this case, irrespective of whether the fluid itself is viscous, its energy--momentum tensor, evaluated at the solution, takes the zeroth-order (perfect) form (\ref{PMP-T0}) because all derivatives of the hydrodynamic variables vanish. On the one hand, this equilibrium situation is easy to handle because the relevant equations are the zeroth-order ones, \eqref{PMP-Euler0}; on the other hand, it does not allow to learn anything about transport properties because the effect of transport is washed out by the geometry itself. If we insist keeping Minkowski as a background, the only way, which would give access to the transport coefficients, is to perturb the fluid away from its global equilibrium configuration.

Although naive, the equilibrium paradigm in Minkowski  has the virtue to suggest an alternative  general method that may fit certain classes of fluids. It indeed raises a less naive question: are there other situations of fluids on gravitational backgrounds, where the hydrodynamic description is also perfect \emph{i.e.} the energy--momentum tensor, in equilibrium, takes the perfect form (\ref{PMP-T0}) solving Eqs. \eqref{PMP-Euler0}? 

As anticipated, we call these special configurations \emph{perfect-equilibrium states}. For these configurations to exist, \emph{all} terms in  \eqref{PMP-T00}, except for the first one, must vanish, either because the  transport coefficients are zero, or because the corresponding tensors vanish kinematically -- requiring in particular a special relationship between the fluid's velocity and the background geometry. It should be stressed that the fluid in perfect equilibrium is \emph{not} perfect -- the equilibrium is. 
 
At this stage of the presentation, the question to answer is whether fluids exist, which can exhibit, on certain backgrounds, perfect-equilibrium configurations. Holography and the methods discussed in Sects. \ref{PMP-sec:anc} and \ref{PMP-sec:sdb} for finding exact bulk solutions provide the tools for this analysis.  The strategy to follow is straightforward:
\begin{itemize}
\item Choose a class of backgrounds possessing a time-like Killing vector $\xi$.
\item Assume perfect equilibrium and show that indeed perfect Euler Eqs. \eqref{PMP-Euler0} are solved for a conformal fluid \emph{i.e.} for a fluid such that $\varepsilon=2p$. A hint for solving them is to impose that the fluid velocity field $\text{u}$ is aligned with $\xi$.
\item Impose the ``self-duality'' condition \eqref{PMP-sdw} and restrict the family of backgrounds at hand. The three-dimensional geometries obtained in that way are called \emph{perfect geometries} because their Cotton--York tensor is of the perfect-fluid form.
\item Use the Fefferman--Graham expansion to reconstruct the four-dimensional bulk geometry, hoping indeed that Eq.  \eqref{PMP-sdw} acts as an integrability condition, allowing for resummation of the series into an exact Einstein space. This is crucial for sustaining the claim that we are describing a holographic  conformal fluid behaving \emph{exactly} as a perfect fluid. 
\end{itemize}

If this procedure goes through with genuinely non-trivial geometries, it enables us to probe transport properties of the holographic fluid despite its global equilibrium state: all transport coefficients coupled to Weyl-covariant, traceless and transverse tensors $\mathcal{T}_{\mu\nu}$ that are non-vanishing  and whose divergence is also non-vanishing, when evaluated in the perfect-equilibrium solution, must be zero. We call such tensors \emph{dangerous tensors}. Listing them requires the knowledge of the specific perfect geometry and of the kinematic configuration of the fluid.\footnote{More data are available on the dangerous tensors in certain classes of geometries in \cite{PMP-Mukhopadhyay:2013gja}.}  Any fluid, which would have non-vanishing corresponding transport coefficient, would not be in equilibrium in the configuration at hand. This may occur for transport coefficients of any order in the expansion of the energy--momentum tensor, as dangerous tensors appear at arbitrarily large derivative order. Therefore the insight gained in this manner on the transport properties of the holographic fluid, concerns usually infinite series of coefficients.
This is a non-trivial piece of information about the conformal fluid at hand, and a statement about the underlying microscopic theory.

There are non-trivial backgrounds (Minkowski space being a trivial example) where no dangerous tensors are present. However, one can also find a large class of backgrounds with a unique time-like Killing vector field, which have infinitely many non-zero dangerous tensors; those allow to probe an infinite number of transport coefficients. It is not clear at present whether all these backgrounds exhaust the perfect geometries. Nevertheless, the question of whether our analysis regarding all possible
transport coefficients is exhaustive or not requires more work. It is clear that further insight on this matter can only be gained by perturbing the perfect-equilibrium state.

 \paragraph{Perfect equilibrium in Papapetrou--Randers backgrounds}

A stationary three-dimensional metric can be written in the generic form ($x=\left(x^1,x^2\right)$
and  $i,j,\ldots=1,2$)
\begin{equation}
\label{PMP-Papa}
\mathrm{d}s^2=B(x)^2\left(-(\mathrm{d}t-b_i(x) \mathrm{d}x^i)^2+a_{ij}(x)\mathrm{d}x^i \mathrm{d}x^j\right),
\end{equation}
where $B, b_i, a_{ij}$ are space-dependent but time-independent functions. These metrics were introduced by Papapetrou in \cite{PMP-Papapetrou}. 
They will be called hereafter \emph{Papapetrou--Randers}  because they are part of an interesting network of relationships involving the Randers form \cite{PMP-Randers}.
These metrics admit a generically \emph{unique} time-like Killing vector,  $\xi\equiv\partial_t$, with norm $\left\| \xi\right\|^2=-B(x)^2$.

At this stage of the analysis, we would like to restrict ourselves to the case where the Killing vector is normalized, \emph{i.e.} where $B$ is constant and can therefore be consistently set to 1. This is a severe limitation, because it excludes equilibrium situations where the temperature or the chemical potential are $x$-dependent.\footnote{Remember that inside a stationary gravitational field, under certain conditions, global thermodynamic equilibrium requires $T\sqrt{-g_{00}}$ 
%and $\mu\sqrt{-g_{00}}$ 
be constant \cite{PMP-LLSP}. Here $\sqrt{-g_{00}}=B$. Holographically, if the rescaling of the boundary metric by $B(x)$ (as in \eqref{PMP-Papa}) is accompanied with an appropriate rescaling of the energy--momentum tensor, the bulk geometry is unaffected, and $B(x)$ is generated by a bulk diffeomorphism.} However, it  illustrates the onset of perfect equilibrium configurations, and allows to establish a wide class of perfect geometries, intimately connected with holography.

In the background \eqref{PMP-Papa} (with $B=1$), the vector $\xi=\partial_t$, satisfies 
\begin{equation}
\nonumber
%\label{PMP-Killing}
\nabla_{(\mu} \xi_{\nu)}=0, \quad \xi_\mu \xi^\mu=-1. 
\end{equation}
We leave as an exercise to show that 
congruences defined by $\xi$ have vanishing acceleration, shear and expansion (see App. \ref{PMP-vfc}), but non-zero vorticity\footnote{Vorticity
 is inherited from the fact that  $\partial_t$ is not hypersurface-orthogonal. For this very same reason, Papapetrou--Randers geometries may in general suffer from global hyperbolicity breakdown. This occurs whenever regions exist, where constant-$t$ surfaces cease being space-like, and potentially exhibit closed time-like curves. All these issues were discussed in detail in \cite{PMP-Leigh:2011au, PMP-LPP2, PMP-NewPaper}.} 
 $\omega = \frac{1}{2}\mathrm{d}\xi \Leftrightarrow 
\omega_{\mu\nu}=\nabla_{\mu}\xi_{\nu}$. Then, it is easy to show that a solution of the perfect Euler equations (\ref{PMP-Euler0}), for a conformal fluid is:
\begin{equation}\label{PMP-Euler0K}
\mathrm{u}= \xi, \quad  \varepsilon = 2p = \, \text{constant}, \quad T = \,\text{constant}, \quad  s = \, \text{constant}. 
\end{equation}
Therefore a fluid in perfect equilibrium will align its velocity field\footnote{One important point to note is that in perfect equilibrium we have no frame ambiguity in defining the velocity field. Since the velocity field is geodesic and is aligned with a Killing vector field of unit norm, it describes a unique local frame where all forces (like those induced by a temperature gradient) vanish.} $\text{u}$ with the vector $\xi=\partial_t$, while thermalize at everywhere-constant $p$ and $T$. Fluid worldlines form a shearless and expansionless geodesic congruence. 

The normalized three-velocity one-form of the fluid at perfect equilibrium is 
\begin{equation}\label{PMP-velform}
\mathrm{u}=-\mathrm{d}t+\mathrm{b},
\end{equation}
where $\mathrm{b}=b_i \mathrm{d}x^i$. We will often write the metric \eqref{PMP-Papa} as 
\begin{equation}\label{PMP-Papas}
\mathrm{d}s^2=-\mathrm{u}^2+\mathrm{d}\ell^2\, \quad \mathrm{d}\ell^2= a_{ij}\, \mathrm{d}x^i\mathrm{d}x^j.
\end{equation}
A conformal fluid  in perfect equilibrium on Papapetrou--Randers backgrounds has the energy--momentum tensor
\begin{equation}
T^{(0)}_{\mu\nu}\,\mathrm{d}x^\mu\mathrm{d}x^\nu=p\left(2\mathrm{u}^2 +\mathrm{d}\ell^2\right)
\label{PMP-T0c}
\end{equation}
with the velocity form being given by \eqref{PMP-velform} and $p$ constant. We will adopt the convention that hatted quantities will be referring to the two-dimensional positive-definite metric $a_{ij}$, therefore $\hat\nabla$ for the covariant derivative and $\hat R_{ij}\,\mathrm{d}x^i\mathrm{d}x^j=\frac{\hat R}{2}\mathrm{d}\ell^2 $ for the Ricci tensor built out of $a_{ij}$. We collect in App. \ref{PMP-app:RP} some useful formulas regarding Papapetrou--Randers backgrounds and the kinematics of fluids at perfect equilibrium. 
 
Let us close this chapter by insisting once more on the meaning of the perfect-equilibrium configuration \eqref{PMP-Euler0K} for a conformal fluid that \emph{is not a priori} perfect. For this configuration to be effectively realized, \emph{all} higher-derivative corrections in \eqref{PMP-T00}  must be absent. It is easy to check that this is indeed the case for the first corrections, which in the $2+1$-dimensional case under consideration read:
\begin{equation}\label{PMP-T1}
T_{(1)}^{\mu \nu}=-2\eta \sigma^{\mu \nu}
      -\zeta_{\mathrm{H}}\eta_{\vphantom{\lambda}}^{\rho\lambda(\mu}u_\rho\sigma_\lambda^{\hphantom{\lambda}\nu)}.
\end{equation}
The first term in (\ref{PMP-T1}) involves the shear viscosity $\eta$, which is a dissipative transport coefficient. The second is present in systems that break parity and involves the non-dissipative rotational-Hall-viscosity coefficient $\zeta_{\mathrm{H}}$. Notice that the bulk-viscosity term $\zeta \Delta^{\mu \nu}\Theta$ or the anomalous term $\tilde{\zeta}\Delta^{\mu\nu}
\eta^{\alpha\beta\gamma}
u_\alpha\nabla_\beta u_\gamma$ cannot appear in a conformal fluid because they are  tracefull, namely for conformal fluids $\zeta =\tilde{\zeta}=0$.  
Since the fluid congruence
is shearless, the first corrections \eqref{PMP-T1} vanish. Demanding that 
higher-order corrections also vanish, on the one hand, sets constraints on the transport coefficients coupled to the dangerous tensors that can be constructed with the vorticity only; on the other hand, it leaves free many other coefficients, which couple to tensors vanishing because of the actual kinematic state of the fluid. If the transport coefficients coupled to the dangerous tensors are non-zero, 
the geodesic fluid congruence with constant temperature is not a solution of the full Euler equations \eqref{PMP-cons}. The resolution of the latter alters the above perfect equilibrium state,
leading in general to $\mathrm{u}=\xi+\delta\mathrm{u}(\mathrm{x})$ and $T=T_0+\delta T(\mathrm{x})$. Such an excursion will be stationary or not depending on whether the non-vanishing corrections to the perfect energy--momentum tensor are non-dissipative or dissipative. 

\subsection{Perfect-Cotton geometries and their bulk ascendents} \label{PMP-pCg}
 
 \paragraph{The strategy}

The analysis presented in Sect. \ref{PMP-FEPR} is useful if there exist conformal fluids, which are indeed in perfect equilibrium on a Papapetrou--Randers background. This is not guaranteed \emph{a priori} since it requires infinite classes of transport coefficients to vanish. Holography provides the appropriate tools for addressing this problem. The strategy has already been described above, and the remaining two steps are the following:
\begin{enumerate}
\item Impose condition \eqref{PMP-sdw} with perfect energy--momentum tensor and hence restrict the Papapetrou--Randers geometries to those which have a Cotton--York tensor of the perfect-fluid form \eqref{PMP-T0c}:
\begin{equation}\label{PMP-perfCott}
C_{\mu\nu}= \frac{c}{2}(3 u_\mu u_\nu + g_{\mu\nu}), 
\end{equation}
where $c$ is a constant with the dimension of an energy density.\footnote{We recall that $\varepsilon$ has dimensions of energy density or equivalently $(\mathrm{length})^{-3}$, therefore the energy--momentum tensor and the Cotton--York tensor have the same natural dimensions. } This form is known in the literature as Petrov class $\text{D}_{\mathrm{t}}$.\footnote{The subscript t stands for \emph{time-like} and refers to the nature of the vector u. For an exhaustive review on Petrov  \& Segre classification of three-dimensional geometries see \cite{PMP-Chow:2009km} (useful references are also \cite{PMP-Guralnik, PMP-Grumiller, PMP-Moutsopoulos:2011ez}).} Notice that the existence of perfect geometries is an issue unrelated to holography. 
\item Sum the Fefferman--Graham series expansion. It turns out that the bulk geometries obtained in this way are exact solutions of Einstein's equations: perfect-Cotton geometries are boundaries of $3+1$-dimensional exact Einstein spaces, and the resulting boundary energy--momentum tensor is also of the perfect-fluid form. This shows that the assumption of perfect equilibrium is well motivated, and the ``self-duality'' condition \eqref{PMP-sdw} does indeed ensure integrability.
\end{enumerate}

 \paragraph{Classification of the perfect Papapetrou--Randers geometries}

Consider a metric of the form \eqref{PMP-Papa} with $B(x)=1$. Requiring its Cotton--York tensor \eqref{PMP-Cgen} to be of the form \eqref{PMP-perfCott} is equivalent to impose the conditions: 
\begin{eqnarray}
\label{PMP-cotequileq1}
\hat\nabla^2 q+q(\delta-q^2)&=&2c,\\ \label{PMP-cotequileq2}
a_{ij}\left(\hat\nabla^2 q +\frac{q}{2}(\delta-q^2)-c\right)
&=& \hat\nabla_i\hat\nabla_j q,\\
\label{PMP-Requil}
\hat R +3 q^2&=&\delta
\end{eqnarray}
with $\delta$ being a constant relating the curvature of the two-dimensional base space, $\hat R$, with the vorticity strength $q$ (see App. \ref{PMP-app:RP} for definitions and formulas).

It is remarkable that perfect-Cotton geometries \emph{always} possess an extra space-like Killing vector. To prove\footnote{I thank Jakob Gath for clarifying this point.} this we rewrite \eqref{PMP-cotequileq1} and \eqref{PMP-cotequileq2}
as
\begin{equation}
\label{PMP-cotequileq3}
\left(\hat\nabla_i\hat \nabla_j-\frac12\,a_{ij} \hat\nabla^2\right)q=0.
\end{equation}
Any two-dimensional metric can be locally written as 
\begin{equation}
\label{PMP-2dcomplex}
\mathrm{d}\ell^2=2\mathrm{e}^{2\Omega(z,\bar z)}\mathrm{d}z\,\mathrm{d}\bar z,
\end{equation}
where $z$ and $\bar z$ are complex-conjugate coordinates. Plugging \eqref{PMP-2dcomplex} in \eqref{PMP-cotequileq3} we find that the non-diagonal 
equations are always satisfied (tracelessness of the Cotton--York tensor), while the diagonal ones read:
\begin{equation}
\nonumber
%\label{PMP-cotequileq4}
\partial_z^2q=2\partial_z \Omega\partial_z q ,\quad \partial_{\bar z}^2q=2\partial_{\bar z} \Omega\partial_{\bar z} q.
\end{equation}
The latter can be integrated to obtain 
\begin{equation}
\label{PMP-cotequileq5}
\partial_zq=\mathrm{e}^{2\Omega-2\bar C(\bar z)},\quad \partial_{\bar z}q=\mathrm{e}^{2\Omega-2C(z)}
\end{equation}
with $C(z)$ an arbitrary holomorphic function and $\bar C(\bar z)$ its complex conjugate. Trading these functions for
\begin{equation}
\nonumber
w(z)=\int \mathrm{e}^{2 C(z)}\mathrm{d}z,\quad \bar w(\bar z)=\int \mathrm{e}^{2 \bar C(\bar z)}\mathrm{d}\bar z,
\end{equation}
and introducing new coordinates $(X,Y)$ as
\begin{equation}
\nonumber
X=w(z)+\bar w(\bar z), \quad Y = i\left(\bar w(\bar z)-w(z)\right), 
\end{equation}
we find using  \eqref{PMP-cotequileq5} that the vorticity strength depends only on $X$: $q=q(X)$.
Hence, \eqref{PMP-2dcomplex} reads: 
\begin{equation}
\nonumber
\text{d}\ell^2=\frac{1}{2}\partial_X q\left(\mathrm{d}X^2+\mathrm{d}Y^2 \right).
\end{equation}
This condition enforces the existence of an extra Killing vector.
Finally we note that \eqref{PMP-Requil} can be obtained by differentiating \eqref{PMP-cotequileq1} with respect to $X$.

The presence of the space-like isometry actually simplifies the perfect-Cotton conditions for Papapetrou--Randers metrics. Without loss of generality, we take the space-like Killing vector to be $\partial_y$ and write the metric as
\begin{equation}
\label{PMP-FH-rot-byG}
\mathrm{d}s^2 = -\left(\mathrm{d}t- b(x)\,\mathrm{d}y\right)^2+ 
\frac{\mathrm{d}x^2}{G(x)}   + G(x) \mathrm{d}y^2.
\end{equation}
Thus 
\begin{equation}
\nonumber
%\label{PMP-gen-q}
q=-\partial_x b,
\end{equation}
and \eqref{PMP-cotequileq1}--\eqref{PMP-Requil} can be solved in full generality. The solution is written in terms of 6 arbitrary parameters $  c_i$:
\begin{eqnarray}
\label{PMP-b}
b(x)&=&  c_0
+   c_1 x+  c_2 x^2, \\
\label{PMP-G}
G(x) &=&   c_5 +   c_4 x+   c_3 x^2+  c_2  x^3 \left( 2  c_1+  c_2 x\right).
\end{eqnarray}
It follows that the vorticity strength takes the linear form
\begin{equation}
\label{PMP-q}
q(x)= -  c_1 - 2   c_2 x,
\end{equation}
and the constants $c$ and $\delta$ are given by:
\begin{eqnarray}\label{PMP-cflat}
c&=& -  c_1^3 +   c_1   c_3 -   c_2   c_4,\\ 
\label{PMP-delflat}
\delta&=& 3  c_1^2-2  c_3 .
\end{eqnarray}
Finally, the Ricci scalar of the two-dimensional base space is given by
\begin{equation}
\nonumber
%\label{PMP-Rbaseequil}
\hat R =-2 \left(  c_3 + 6  c_2 x(  c_1 +   c_2 x) \right), 
\end{equation}
and using \eqref{PMP-Rsgen} one can easily find the form of the three-dimensional scalar curvature as well. 
Not all the six parameters $  c_i$ correspond to physical quantities: some of them can be just reabsorbed in a change of coordinates. In particular, we set here $  c_0 = 0$ by performing the diffeomorphism $t \to t + p\, y$, with constant $p$, which does not change the form of the metric. 

 \paragraph{The bulk duals of the perfect geometries}

At this stage, the reader may wonder what the interpretation of the parameters $  c_i$ is. It is more convenient to answer that question after unravelling the Einstein metrics that fit the boundary data \eqref{PMP-T0c}, and \eqref{PMP-FH-rot-byG} (with \eqref{PMP-b} and \eqref{PMP-G}). As already advertised, with these boundary data, the Fefferman--Graham series  is resummable because \eqref{PMP-T0c} and \eqref{PMP-perfCott} satisfy the ``self-duality'' condition \eqref{PMP-sdw} with $w=\nicefrac{-c}{\varepsilon}$. The resulting exact Einstein space reads, in Eddington--Finkelstein coordinates (where $g_{rr}=0$ and $g_{r\mu} = - u_\mu$):
\begin{eqnarray}\label{PMP-4d.CpropT}
\mathrm{d}s^2 &=& - 2 \mathrm{u} \left( \mathrm{d}r- \frac{1}{2k^2} G(x) \partial_x q  \, \mathrm{d}y\right) + \rho^2 k^2\mathrm{d}\ell^2 \nonumber \\
 &&- 
\left(r^2k^2+\frac{\delta}{2k^2}-\frac{q^2}{4k^2}
-\frac{1}{\rho^2}\left(2Mr+ \frac{ q c}{2 k^6}\right) \right)\mathrm{u}^2 
\end{eqnarray}
with
\begin{eqnarray}
\mathrm{u} &=& -\mathrm{d}t + b\, \mathrm{d}y
,\\
\label{PMP-rho}
\rho^2 &=& r^2 + \frac{q^2}{4k^4}.
\end{eqnarray}
The various quantities appearing in \eqref{PMP-4d.CpropT}--\eqref{PMP-rho}, $b(x)$, $G(x)$, $q(x)$, $c$ and $\delta$, are reported in Eqs. \eqref{PMP-b}--\eqref{PMP-delflat}. Notice also that a coordinate transformation is needed in order to recast \eqref{PMP-4d.CpropT} in Boyer--Lindqvist coordinates, and a further one to move to the canonical Fefferman--Graham frame \eqref{PMP-FGf}. Details can be found in \cite{PMP-Mukhopadhyay:2013gja}, which we will not present here because they lie beyond the main scope of these lectures. Even though $r$ is not the Fefferman--Graham radial coordinate, in the limit  $r\rightarrow \infty$, they both coincide. It is easy then to see that the boundary geometry is indeed the  stationary Papapetrou--Randers metric \eqref{PMP-Papas}, \eqref{PMP-FH-rot-byG}, and that the boundary energy--momentum tensor is of the perfect-fluid form with
\begin{equation}
\nonumber
%\label{PMP-eden}
\varepsilon = \frac{Mk^2}{4\pi G_{\mathrm{N}}}.
\end{equation}

Upon performing coordinate transformations and parameter redefinitions, one can show that for $c_4\neq 0$, the bulk metrics at hand belong to the general class of Pleba\~{n}ski--Demia\`{n}ski type $\text{D}$, analyzed in \cite{PMP-PD}. For vanishing $c_4$, depending on the other parameters, one finds the flat-horizon solution of \cite{PMP-Ortin}, or the rotating topological black hole of \cite{PMP-Klemm:1997ea}, or a set of metrics, which were found (but still not fully studied) in \cite{PMP-Mukhopadhyay:2013gja}. All these solutions are AdS 
black holes, which have mass $M$, nut charge $n$ and angular velocity $a$. The acceleration parameter, present in  Pleba\~{n}ski--Demia\`{n}ski \cite{PMP-PD} is missing here. Actually, this parameter  is an obstruction to  perfect-Cotton boundary (\emph{i.e.} to $\text{D}_{\text{t}}$ Petrov--Segre class), and this is why it does not appear in our classification (see also \cite{PMP-Klemm:2014nka}). 

For all these metrics, the horizon is spherical, flat or hyperbolic.\footnote{This is a local property. In the flat or hyperbolic cases, a quotient by a discrete subgroup of the isometry group is possible and allows to reshape the global structure, making the horizon compact without conical singularities (a two-torus for example).} The isometry group contains at least the time-like Killing vector $\partial_t$ and the space-like Killing vector $\partial_y$. In the absence of rotation, two extra Killing fields appear,  which together with $\partial_y$ generate $SU(2)$, Heisenberg or $SL(2, \mathbb{R})$. The bulk metric is then a foliation over Bianchi IX, II or VIII homogeneous geometries.

From the explicit form of the bulk space--time metric \eqref{PMP-4d.CpropT}, we observe that it can have a curvature singularity when $\rho^2 = 0$. The locus of this singularity will then be at 
\begin{equation}
\nonumber
%\label{PMP-sing}
r = 0, \quad q(x) = 0.
\end{equation}
It also has an ergosphere, where  the Killing vector $\partial_t$ becomes null,\footnote{The  Killing vector $\partial_t$ is time-like and normalized at the boundary, where it coincides with the velocity field of the fluid, but its norm gets altered along the holographic coordinate, towards the horizon.} at $r(x)$ solution of 
\begin{equation}
\nonumber
r^2k^2+\frac{\delta}{2k^2}-\frac{q^2}{4k^2}
-\frac{1}{\rho^2}\left(2Mr+ \frac{ q c}{2 k^6}\right)= 0.
\end{equation}

We will not pursue any further this discussion on the bulk geometries. A thorough analysis of horizons, singularities or closed time-like curves can be found in the already quoted literature. A last comment concerning these black holes should however be made in relation with their symmetries: they are stationary and possess at least an additional spatial isometry. This is a consequence of the perfect-Cotton structure of their boundary, and this is consistent with the rigidity theorem in $3+1$ dimensions, which requires all stationary black hole solutions in flat space--time to have an axial symmetry. However, as far as we are aware, it is not known if this theorem is valid for $3+1$-dimensional asymptotically AdS stationary black holes. The above analysis  appears thus as an indirect and somehow unexpected hint in favor of the rigidity theorem beyond asymptotically flat space--times.

Let us end this paragraph with an example, which generalizes (in Lorentzian signature) the case \eqref{PMP-hyptaubnut} presented in Sect. \ref{PMP-sec:ce}: the AdS Kerr--Taub--NUT with spherical horizon. In Boyer--Lindqvist coordinates this reads:
 \begin{equation}
 \label{PMP-AdSKTN}
\mathrm d s^2 = \frac{\rho^2}{\Delta_r}\mathrm{d}r^2 
- \frac{\Delta_r}{\rho^2}\left(\mathrm{d}t + \beta\mathrm{d}{ \varphi} \right)^2
+\frac{\rho^2}{\Delta_\vartheta}\mathrm{d}\vartheta^2 +  \frac{\sin^2\vartheta\Delta_\vartheta}{\rho^2}\left(a \mathrm{d}t +  \alpha \mathrm{d}{\varphi} \right)^2
\end{equation}
with
\begin{eqnarray}
\nonumber
\rho^2 &=& r^2  + (n- a \cos\vartheta)^2, \\
\Delta_r &=&k^2  r^4 +r^2 (1 + k^2 a^2  + 6 k^2 n^2) -2 M r + (a^2 - n^2)  (1 + 3 k^2 n^2 ) , 
\nonumber
\\
\Delta_\vartheta &=&  1 +k^2 a \cos\vartheta(4n - a \cos\vartheta) 
\nonumber%,\\
%\beta &=&  -\frac{2 (a - 2 n + a \cos\vartheta)}{\Xi}\sin^2\nicefrac{\vartheta}{2},\\
%\alpha &=& -\frac{r^2+(n-a)^2}{\Xi}, \\
%\Xi &=& 1- k^2 a^2 .
 \end{eqnarray}
 and 
 \begin{eqnarray}
%\rho^2 &=& r^2  + (n- a \cos\vartheta)^2, \\
%\Delta_r &=&k^2  r^4 +r^2 (1 + k^2 a^2  + 6 k^2 n^2) -2 M r + (a^2 - n^2)  (1 + 3 k^2 n^2 ) , \\
%\Delta_\vartheta &=&  1 +k^2 a \cos\vartheta(4n - a \cos\vartheta), \\
\nonumber
\beta &=&  -\frac{2 (a - 2 n + a \cos\vartheta)}{\Xi}\sin^2\nicefrac{\vartheta}{2},\\
\nonumber
\alpha &=& -\frac{r^2+(n-a)^2}{\Xi}, \\
\nonumber
\Xi &=& 1- k^2 a^2 .
  \end{eqnarray}
 
 \paragraph{Back to the boundaries and transport properties}

The boundary physics depends on the subset of those parameters among the $c_i$s, which are non-trivial.  The boundary metric is in general a function of two parameters, $n$ and $a$, whereas $M$ appears in the boundary energy--momentum tensor. The bulk isometry group is conserved. Thus, in the absence of rotation parameter $a=0$, the boundary is a homogeneous and stationary space--time: squashed $S^3$ (including e.g. G\"odel space), squashed Heisenberg or squashed $\text{AdS}_3$. The fluid undergoes a homogeneous rotation (\emph{i.e.} without center, \emph{monopolar}) with constant vorticity strength $q$. 

For non-vanishing $a$, the boundary space--time is stationary but has only spatial axial symmetry. The vorticity is a superposition of a \emph{monopole} and a \emph{dipole}, and the fluid has now a cyclonic rotation around the poles on top of the uncentered one. 

We give for illustration the boundary metric of the Kerr--Taub--NUT space--time with spherical horizon \eqref{PMP-AdSKTN}:
\begin{equation}
 \label{PMP-AdSKTNb}
\text{d}s^2_{\text{bry.}}= 
-\left(\mathrm{d}t + \beta\mathrm{d}{ \varphi} \right)^2
+\frac{1}{k^2\Delta_\vartheta}\left(\mathrm{d}\vartheta^2 +\frac{\Delta_\vartheta^2}{\Xi^2}\sin^2\vartheta \, \mathrm{d}{\varphi}^2\right). 
\end{equation}
For vanishing $a$, $\text{d}\ell^2$ is an ordinary  two-sphere and $\text{b}=-\beta \text{d}\varphi$ is a Dirac-monopole-like potential. Switching-on $a$ deforms axially the base space $\text{d}\ell^2$, while it adds a dipole contribution to $\text{b}$. From the perspective of transport in holographic fluids, the purpose is to list the dangerous tensors carried by this kind of boundaries. The more tensors we have, the more information we gain on vanishing transport coefficients: since the energy--momentum tensor that emerges holographically is perfect, any transport coefficient coupled to a dangerous tensor is necessarily zero.

For the boundary metric \eqref{PMP-AdSKTNb}, the vorticity strength, the Cotton prefactor and the scalar curvature read:
\begin{eqnarray}
\nonumber
q&=&2k^2(n-a\cos \vartheta),\\
\nonumber
c&=&2k^4 n\left(1+k^2\left(4n^2-a^2\right)\right),\\
\nonumber
R&=&2k^2\left(1+k^2n^2+10 k^2 n a \cos\vartheta+
k^2 a^2\left(1-5\cos^2 \vartheta\right)\right).
 \end{eqnarray}
We observe that, on the one hand, the nut charge $n$ is responsible for the $2+1$-dimensional boundary not being conformally flat. The ordinary rotation parameter $a$, on the other hand, introduces a $\vartheta$-dependence in $q$ and $R$. This betrays the breaking of homogeneity due to $a$: when $a$ vanishes, the boundary is an squashed $S^3$ with $SU(2)\times \mathbb{R}$ isometry, which is a homogeneous space--time, and all of its scalars are constants.\footnote{This family includes G\"odel space--time (see \cite{PMP-rayPRD80, PMP-rebPRD83} for more information). The important issue of closed time-like curves emerges as a consequence of the lack of global hyperbolicity. This was discussed in Refs. \cite{PMP-Leigh:2011au, PMP-LPP2, PMP-NewPaper}, in relation with holographic fluids. When the bulk geometry has hyperbolic horizon, this caveat can be circumvented.}

Coming back to the discussion on the dangerous tensors, we expect them to be more numerous when less symmetry is present. Indeed, for vanishing $a$, all scalars are constant and both the Riemann and the Cotton are combinations of $u_\mu u_\nu$  and $g_{\mu \nu}$ with constant coefficients. Any covariant derivative acting on those will be algebrised in a similar fashion. Thus 
\begin{itemize}
\item all hydrodynamic scalars are constants,
\item all hydrodynamic vectors are of the form $A u_\mu$ with constant $A$, and
\item all hydrodynamic tensors are of the form $B u_\mu u_\nu + C g_{\mu\nu} $ with constant $B$ and $C$.
\end{itemize}
Hence there exists no traceless transverse tensor that can correct the hydrodynamic energy--momentum tensor in perfect equilibrium. In other words, there is no dangerous tensor. Therefore, in the case of monopolar geometries, the symmetry is too rich and  in such a highly symmetric kinematical configuration, the fluid dynamics cannot be sensitive to any dissipative or non-dissipative coefficient. 
As soon as a dipole component is added ($a \neq 0$), a space-dependence emerges in the various scalars and tensors, and infinitely many dangerous tensors appear, which provide valuable information on the vanishing transport coefficients of the holographic fluid. 

The above discussion provides a guide for the subsequent analysis. To have access to more transport coefficients, we must perturb the geometry in a way organized e.g. as a multipolar expansion: the higher the multipole in the geometry, the richer the spectrum of transport coefficients that can contribute, if non-vanishing, to the state of the fluid. No exact Einstein spaces are however available beyond dipole configuration (Kerr).\footnote{In 1919, Weyl exhibited multipolar Ricci-flat solutions, which do not seem extendible to the Einstein case (see \cite{PMP-SKMHH} for details).} Thus, this programme lies outside of the present framework, as it requires to 
work with perturbed bulk Einstein spaces, and handle fluid perturbations potentially bringing the fluid away from perfect equilibrium.

\section{Monopolar boundaries and  topologically massive gravity}

Monopolar geometries have been mentioned in Sect. \ref{PMP-pCg} around the example \eqref{PMP-AdSKTNb}, which appears as the boundary of Taub--NUT Schwarzschild  AdS black hole with spherical horizon.
This terminology is justified by the fact that the vorticity strength $q$ is constant (like the strength of the magnetic field on a sphere surrounding a Dirac monopole).  Within the perfect-Cotton Papapetrou--Randers geometries \eqref{PMP-FH-rot-byG}, there is a whole class of monopolar boundaries, obtained by setting $c_2=0$ in \eqref{PMP-q}. With constant $q$, using the general equations \eqref{PMP-Rsgen} and \eqref{PMP-Rgen} for Papapetrou--Randers, as well as \eqref{PMP-perfCott}--\eqref{PMP-Requil} for perfect-Cotton geometries, we find:
\begin{eqnarray}
\nonumber
%\label{PMP-RsPRPC}
R &=& \delta-\frac{5q^2}{2},\\
\nonumber
%\label{PMP-RRRPC}
R_{\mu\nu}\,\mathrm{d}x^\mu\mathrm{d}x^\nu&=&
 \frac{\delta - q^2}{2}\mathrm{u}^2 +\left(\frac{\delta}{2}-q^2\right)\mathrm{d}s^2,\\
\nonumber
%\label{PMP-CRRPC}
C_{\mu\nu}\,\mathrm{d}x^\mu\mathrm{d}x^\nu&=& \frac{q}{4}\left(\delta-q^2\right)
\left(3\text{u}^2+\text{d}s^2\right).
\end{eqnarray}
These expressions can be combined into
\begin{equation}
\label{PMP-TMGeq}
R_{\mu \nu} -\frac{R}{2}g_{\mu \nu}+\lambda g_{\mu \nu}=\frac{1}{\mu}
C_{\mu \nu}
\end{equation}
with 
\begin{equation}
\nonumber
%\label{PMP-TMGmu}
\lambda =\frac{\delta}{6}-\frac{5q^2}{12}, \quad \mu = \frac{3q}{2}.
\end{equation}

Expression \eqref{PMP-TMGeq} shows that monopolar geometries solve the topologically massive gravity equations \cite{PMP-cs} for appropriate constants $\lambda$ and $\mu$. This is not surprising, as it is a known fact that, for example, squashed anti-de-Sitter or squashed three-spheres solve topologically massive gravity  equations \cite{PMP-Chow:2009km, PMP-Guralnik, PMP-Grumiller, PMP-Moutsopoulos:2011ez}. However, what is worth stressing here is that reversing the argument and requiring a generic Papapetrou--Randers background \eqref{PMP-Papa}  to solve \eqref{PMP-TMGeq} leads \emph{necessarily} to a monopolar geometry. We leave as an exercise to set that result.\footnote{Use the expression for the Ricci tensor for Papapetrou--Randers geometries \eqref{PMP-Rgen}, impose tracelessness and extract $\lambda$. Then use \eqref{PMP-Cgen} and \eqref{PMP-G} and conclude that $q$ must be constant and related to $\mu$. Combine these results and reach the conclusion that all solutions are fibrations over a  two-dimensional space with metric $\mathrm{d}\ell^2$ of constant curvature $\hat R= 6 \lambda-\nicefrac{2\mu^2}{9}$. They are thus homogeneous spaces of either positive ($S^2$), null ($\mathbb{R}^2$) or negative curvature ($H_2$).}
 
As already advertised, the topological mass term (resulting from the Chern--Simons action in \eqref{PMP-eq:sdca}) appears explicitly, in the cases under consideration, as part of topologically massive gravity equations.  The reader might be puzzled by this connection. The $2+1$-dimensional geometries analyzed here are not supposed to carry any gravity degree of freedom since they are ultimately designed to serve as holographic boundaries. Hence, the emergence of topologically massive gravity  should not \emph{a priori} be considered as a sign of dynamics. Nevertheless, as for the general ``self-dual'' case (Eqs. \eqref{PMP-sdw} obtained by varying \eqref{PMP-eq:sdca}), we should leave open the option of introducing some topologically massive graviton dynamics on the boundary. This approach should not be confused with that of some recent works \cite{PMP-Anninos:2008fx, PMP-Anninos:2011vd}, where topologically massive gravity and its homogeneous solutions  play the role of \emph{bulk} geometries. Investigating the interplay between these two viewpoints might be of some relevance.

\section{Outlook}

Modified versions of Einstein's gravity are of interest primarily in cosmology. The aim of the present lectures is to set a bridge with a somewhat less expected area of applications, namely holography. Prior to holography we actually find, in four-dimensional Euclidean framework, quaternionic spaces. These, from the Fefferman--Graham viewpoint, require a boundary condition, which is obtained holographically as the extremization of 
\begin{equation}
\label{PMP-eq:mcs}
S=S_{\text{holographic matter}} +S_{\text{Chern--Simons}} .
\end{equation}
Assuming homogeneity for the boundary metric, further restricts \eqref{PMP-eq:mcs} to the topologically massive gravity action, as shown in the last paragraph of these notes. Although, at this stage, only the extremum of this action is relevant, investigating boundary graviton dynamics in holographic set-ups might prove interesting in the future. 

Translating the bulk Weyl self-duality condition into boundary data opens up the possibility to make it applicable for Lorentzian-signature bulk and boundary geometries. This sort of integrability requirement is not necessary, however, and many Einstein spaces exist, which do not satisfy \eqref{PMP-sdw}.\footnote{As usual with instantons, self-duality selects ground states, but exact excited states can also exist.}  
Investigating further the relationships amongst the boundary energy--momentum tensor and the boundary Cotton tensor may be instructive in the case of exact Einstein spaces, which fall outside of the class studied here.\footnote{Recently this was discussed for a non-stationary solution of Einstein's equations \cite{PMP-deFreitas:2014lia}.\label{ns}} This could be useful both for understanding the underlying gravitational structure and for studying transport properties in conformal holographic fluids.

Besides potential generalizations of \eqref{PMP-sdw}, appears also here the issue of the form 
of the boundary metric and of the energy--momentum tensor. Our analysis has been limited  to (\romannumeral1) stationary Papapetrou--Randers boundary geometries \eqref{PMP-Papa} with $B=1$, and (\romannumeral2) perfect-fluid-like boundary energy--momentum tensors. These options make operational the determination of vanishing transport coefficients by imposing perfect equilibrium, which turns out to exist holographically. We may however scan more general situations as many more exact Einstein spaces exist that deserve to be analyzed. We have already quoted in Sect. \ref{PMP-pCg} the Pleba\~{n}ski--Demia\`{n}ski Einstein stationary solutions \cite{PMP-PD}, for which the acceleration 
parameter is a source of deviation from the perfect-Cotton boundary geometry. Non-stationary spaces provide equally interesting laboratories for further investigation (see footnote \ref{ns}). Finally, 
on the Euclidean side, a great deal of techniques (isomonodromic deformations, twistors, \dots) have been developed for finding the families of quaternionic spaces quoted in Refs. \cite{PMP-P85, PMP-P86, PMP-przanowski90, PMP-Tod94, PMP-Hitchin95, PMP-CP00, PMP-LeBrun-88} (see also \cite{PMP-Petropoulos:2012ne} for a review). Among these, the Calderbank--Pedersen two-Killing family \cite{PMP-CP02} is particularly interesting, because it includes the Euclidean Weyl-self-dual version\footnote{In this case, \eqref{PMP-sd-QTN} is traded for $M= n\left(1-k^2\left(4n^2-a^2\right)\right)$ (see also \cite{Behrndt:2002xm}).} of the Kerr--Taub--NUT \eqref{PMP-AdSKTN}. Since this family contains more self-dual metrics than our exhaustive analysis of Sect. \ref{PMP-pCg} has revealed, these metrics must necessarily lead to a non-perfect boundary energy--momentum tensor, potentially combined with a Papapetrou--Randers boundary geometry with non-constant $B$. Although this discussion is valid in the Euclidean and not all Euclidean solutions admit a real-time continuation, it should help clarifying the landscape of self-duality holographic properties, and possibly  be useful for Lorentzian extensions.

Last, but very intriguing, comes the limitation in the dimension. We have been analyzing four dimensional bulk geometries because our guideline was self-duality, which indeed exists in this (Euclidean) framework. It can however be generalized in eight-dimensional spaces. There, it is known that the octonionic symbols $\Psi^{ABCD}$  allow to define a duality map:  $\tilde{\mathcal{R}}^{AB}=\Psi^{ABCD}\mathcal{R}_{CD}$. Reducing the Riemann two-form $\mathcal{R}_{AB}$, which belongs to the  $\mathbf{28}$ of $SO(8)$, with respect to $\text{Spin}_7\subset SO(8)$ leads to a self-dual component 
$\mathcal{S}_{\mathbf{21}}$ and an anti-self-dual one $\mathcal{A}_{\mathbf{7}}$. Equations \eqref{PMP-scurv} and \eqref{PMP-acurv} are now traded for 
   \begin{eqnarray}
   \nonumber
         \mathcal{S}_{\mathbf{21}}&=&W^{\mathbf{168}} \phi_{\mathbf{21}}+s\phi_{\mathbf{21}}+W^{\mathbf{105}} \chi_{\mathbf{7}},
         \\  
          \nonumber
   \mathcal{A}_{\mathbf{7}}&=&W^{\mathbf{27}} \chi_{\mathbf{7}} +s \chi_{\mathbf{7}} +S^{\mathbf{35}}\phi_{\mathbf{21}},
        \end{eqnarray}
where the singlet $s$ is the scalar curvature, $S^{\mathbf{35}}$ is the traceless Ricci, and the $W^{\mathbf{I}}$ are the three irreducible components of the Weyl tensor.
Riemann self-dual gravitational instantons, obtained by setting $ \mathcal{A}_{\mathbf{7}}=0$, are known to exist \cite{PMP-Acharya:1996tw, PMP-Floratos:1998ba, PMP-Bakas:1998rt, PMP-Bilal:2001an, PMP-Hernandez:2002fb}. Those are Ricci flat. The question is still open to find Weyl self-dual Einstein spaces, by demanding $S^{\mathbf{35}}=0$ and $W^{\mathbf{27}} =0$. From the boundary perspective,  $W^{\mathbf{27}} =0$ could be interpreted as the extremization requirement for \eqref{PMP-eq:mcs} with respect to the  seven-dimensional boundary metric, the Chern--Simons being now the seven-dimensional one \cite{PMP-Zanelli}.
\section*{Acknowledgements}
The present notes will appear in the proceedings of the 7th Aegean summer school \textsl{Beyond Einstein's theory of gravity}, held in Paros, Greece, September 23 -- 28, 2013. I wish to thank the organizers of this school,  where the lectures were delivered. The material presented here is borrowed from recent or on-going works realized in collaboration with  M. Caldarelli, C. Charmousis, J.--P. Derendinger, J. Gath, R. Leigh, A. Mukhopadhyay, A. Petkou, V. Pozzoli, K. Sfetsos, K. Siampos and P. Vanhove. I also benefited from interesting discussions with I. Bakas, D. Klemm, N. Obers and Ph. Spindel.  The feedback from the Southampton University group was also valuable during a recent presentation of this work in their seminar. This research was supported by the LABEX P2IO, the ANR contract  05-BLAN-NT09-573739, the ERC Advanced Grant  226371. 

%
%\appendix
%\section*{Appendix}
%\addcontentsline{toc}{section}{Appendix}
%
%
\appendix

\section{On vector-field congruences}\label{PMP-vfc}

We consider a manifold endowed with a space--time metric of the generic form
\begin{equation}
\nonumber
%\label{PMP-Dmet}
\mathrm{d}s^2 =g_{\mu\nu}\mathrm{d}\mathrm{x}^\mu \mathrm{d}\mathrm{x}^\nu
= \eta_{\mu\nu}E^\mu E^\nu
\end{equation}
(to avoid inflation of indices we do not distinguish between flat and curved ones).
%We will use $a,b,c,\ldots =0,1,\ldots, D-1$ for transverse Lorentz indices along with $\alpha,\beta,\gamma=1,\ldots, D-1$. Coordinate indices will be denoted $\mu,\nu,\rho, \ldots$ for space--time $\mathrm{x}\equiv (t,x)$ and  $i,j,k, \ldots$ for spatial $x$ directions. 
Consider now an arbitrary time-like vector field $\mathrm{u}$, normalised as  $u^\mu u_\mu=-1$,  later identified with the fluid velocity. Its integral curves define a congruence which is characterised by its acceleration, shear, expansion and vorticity (see e.g. \cite{PMP-Ehlers:1993gf,PMP-vanElst:1996dr}):
\begin{equation}
\nonumber
%\label{PMP-def1}
\nabla_{\mu} u_\nu=-u_\mu a_\nu +\frac{1}{D-1}\Theta \Delta_{\mu\nu}+\sigma_{\mu\nu} +\omega_{\mu\nu}
\end{equation}
with\footnote{Our conventions are: $A_{(\mu\nu)}=\nicefrac{1}{2}\left(A_{\mu\nu}+A_{\nu\mu}\right)$ and $A_{[\mu\nu]}=\nicefrac{1}{2}\left(A_{\mu\nu}-A_{\nu\mu}\right)$.}
\begin{eqnarray}
a_\mu&=&u^\nu\nabla_\nu u_\mu, \quad
\Theta=\nabla_\mu u^\mu, 
%\label{PMP-def21}
\nonumber
\\
\sigma_{\mu\nu }&=&\frac{1}{2} \Delta_\mu^{\hphantom{\mu}\rho } \Delta_\nu ^{\hphantom{\nu }\sigma}\left(
\nabla_\rho  u_\sigma +\nabla_\sigma  u_\rho 
\right)-\frac{1}{D-1} \Delta_{\mu\nu }\Delta^{\rho \sigma } \nabla_\rho  u_\sigma  
%\label{PMP-def22}
\nonumber
\\
&=& \nabla_{(\mu} u_{\nu )} + a_{(\mu} u_{\nu )} -\frac{1}{D-1} \Delta_{\mu\nu } \nabla_\rho  u^\rho  ,
%\label{PMP-def23}
\nonumber
\\
\omega_{\mu\nu }&=&\frac{1}{2} \Delta_\mu^{\hphantom{\mu}\rho } \Delta_\nu ^{\hphantom{\nu }\sigma }\left(
\nabla_\rho  u_\sigma -\nabla_\sigma  u_\rho 
\right)= \nabla_{[\mu} u_{\nu ]} + u_{[\mu} a_{\nu ]}.
%\label{PMP-def24}
\nonumber
\end{eqnarray}
The latter allows to define the vorticity form as
\begin{equation}\label{PMP-def3}
2\omega=\omega_{\mu\nu }\, \mathrm{d}\mathrm{x}^\mu\wedge\mathrm{d}\mathrm{x}^\nu  =\mathrm{d}\mathrm{u} +
\mathrm{u} \wedge\mathrm{a} .
\end{equation}
The time-like vector field $\mathrm{u}$  has been used to decompose any tensor field on the manifold in transverse and longitudinal components.  The decomposition is performed by introducing the longitudinal and transverse projectors:
\begin{equation}
\label{PMP-proj}
U^\mu_{\hphantom{\mu}\nu } = - u^\mu u_\nu , \quad \Delta^\mu_{\hphantom{\mu}\nu } =  u^\mu u_\nu  + \delta^\mu_{\nu },
\end{equation}
where $\Delta_{\mu \nu }$ is also the induced metric on the surface orthogonal  to $\mathrm{u}$. The projectors satisfy the usual identities:
\begin{equation}
\nonumber
U^\mu_{\hphantom{\mu}\rho } U^\rho _{\hphantom{\rho }\nu } = U^\mu_{\hphantom{\mu}\nu },\quad U^\mu_{\hphantom{\mu}\rho } \Delta^\rho _{\hphantom{\rho }\nu }  =   0 , \quad \Delta^\mu_{\hphantom{\mu}\rho } \Delta^\rho _{\hphantom{\rho }\nu }  =   \Delta^\mu_{\hphantom{\mu}\nu } ,\quad U^\mu_{\hphantom{\mu}\mu}=1, \quad \Delta^\mu_{\hphantom{\mu}\mu}=D-1,
\end{equation}
and similarly:
\begin{equation}
\nonumber
u^\mu a_\mu=0, \quad u^\mu \sigma_{\mu\nu }=0,\quad u^\mu \omega_{\mu\nu }=0, \quad u^\mu \nabla_\nu  u_\mu=0, \quad \Delta^\rho _{\hphantom{\rho }\mu} \nabla_\nu  u_\rho  =\nabla_\nu  u_\mu.
\end{equation}

\section{Papapetrou--Randers backgrounds and aligned fluids}\label{PMP-app:RP}

In this appendix, we collect a number of useful expressions for stationary Papapetrou--Randers three-dimensional geometries \eqref{PMP-Papa} with $B=1$, and for fluids in perfect equilibrium on these backgrounds. The latter follow geodesic congruences, aligned with the normalized Killing vector $\partial_t$, with velocity one-form given in \eqref{PMP-velform}. 

We introduce the inverse two-dimensional metric $a^{ij}$, and $ b^i$
such that
\begin{equation}
\nonumber
a^{ij}a_{jk}= \delta^i_k,\quad  b^i=a^{ij}b_j.
\end{equation}
The three-dimensional metric components read:
\begin{equation}
\nonumber
g_{00}=-1,\quad g_{0i}= b_i, \quad
g_{ij}=a_{ij}-b_ib_j,
\end{equation}
and those of the inverse metric:
\begin{equation}
\nonumber
g^{00}=a^{ij}b_i b_j-1,\quad g^{0i}=b^i, \quad
g^{ij}=a^{ij}.
\end{equation}
Finally, 
\begin{equation}
\nonumber
\sqrt{\vert g\vert}=\sqrt{a},
\end{equation}
where $a$ is the determinant of the symmetric matrix with entries $a_{ij}$.

Using \eqref{PMP-velform} and \eqref{PMP-def3} we find that the vorticity of the aligned fluid can be written as the following two-form (the acceleration term is absent here)
\begin{equation}
\nonumber
\omega = \frac{1}{2}\omega_{\mu\nu}\mathrm{d}x^\mu\wedge\mathrm{d}x^\nu=\frac{1}{2}\mathrm{d}\mathrm{b}.
\end{equation}
The Hodge-dual of $\omega_{\mu\nu}$  is
\begin{equation}
\nonumber
\psi^\mu = \eta^{\mu\nu\rho}\omega_{\nu\rho}
\Leftrightarrow
\omega_{\nu\rho} =-\frac{1}{2}\eta_{\nu\rho \mu}\psi^\mu.
\end{equation}
In $2+1$ dimensions it is aligned with the velocity field:
\begin{equation}
\nonumber
\psi^\mu =q u^\mu,
\end{equation}
where, in our set-up, 
\begin{equation}
\nonumber
\label{PMP-qgen}
q(x)=-\frac{\epsilon^{ij}\partial_ib_j}{\sqrt{a}}.
\end{equation}
It is a static scalar field that we call the \emph{vorticity strength}, carrying dimensions of inverse length. Together  with $\hat R(x)$ -- the curvature of the two-dimensional metric $\text{d}\ell^2$ introduced in \eqref{PMP-Papas}, the above scalar carries all relevant information for the curvature of the Papapetrou--Randers geometry. We quote for latter use the three-dimensional curvature scalar:
\begin{equation}
R = \hat R+\frac{q^2}{2},
\label{PMP-Rsgen}
\end{equation}
the three-dimensional Ricci tensor 
\begin{equation}
R_{\mu\nu}\,\mathrm{d}x^\mu\mathrm{d}x^\nu=
 \frac{q^2}{2}\mathrm{u}^2 +\frac{\hat R+q^2}{2}\mathrm{d}\ell^2
-\mathrm{u}\, \mathrm{d}x^\rho u^\sigma \eta_{\rho\sigma\mu} \nabla^\mu q,
\label{PMP-Rgen}
\end{equation}
as well as the  three-dimensional Cotton--York  tensor:
\begin{eqnarray}
C_{\mu\nu}\,\mathrm{d}x^\mu\mathrm{d}x^\nu&=& \frac{1}{2}\left(\hat\nabla^2 q+\frac{q}{2}(\hat R+2q^2)\right)
\left(2\mathrm{u}^2 +\mathrm{d}\ell^2\right)\nonumber
\\ \nonumber
&&-\frac{1}{2}\left(\hat\nabla_i\hat\nabla_j q\,  \mathrm{d}x^i \mathrm{d}x^j  +\hat\nabla^2 q
\, \mathrm{u}^2\right) \\
&&-\frac{\mathrm{u}}{2}\mathrm{d}x^\rho u^\sigma \eta_{\rho\sigma\mu} \nabla^\mu (\hat R +3 q^2).
\label{PMP-Cgen}
\end{eqnarray}

\end{document}